\documentclass[aps,onecolumn,superscriptaddress,prx,preprint,longbibliography]{revtex4-2} 
\usepackage{graphicx}
\usepackage{bigints}
\usepackage{dcolumn}
\usepackage{bm}
\usepackage{color}
\usepackage[tight]{subfigure}
\usepackage{array}
\usepackage{titlesec}
\usepackage{filecontents}
\usepackage{xcolor}
\definecolor{darkblue}{rgb}{0.,0.,0.4}
\definecolor{darkred}{rgb}{0.5,0.,0.}
\definecolor{BlueViolet}{RGB}{138,43,226}
\definecolor{SkyBlue}{RGB}{30,144,255}
\definecolor{DarkGreen}{RGB}{0,100,0}
\usepackage[pdftex,colorlinks=true,linkcolor=darkblue,citecolor=red,urlcolor=darkred]{hyperref}
\usepackage{float}
\usepackage{physics}
\usepackage{cancel}
\usepackage[version=4]{mhchem}
\usepackage{soul}
\usepackage{bigints}
\newcommand{\bk}{\mathbf{k}}
\newcommand{\bp}{\mathbf{p}}
\newcommand{\bq}{\mathbf{q}}
\newcommand{\br}{\mathbf{r}}

\def \nn{\nonumber \\}

\newcommand{\bea}{\begin{eqnarray}}
\newcommand{\eea}{\end{eqnarray}}
\newcommand{\beq}{\begin{equation}}
\newcommand{\eeq}{\end{equation}}

\DeclareUnicodeCharacter{2212}{-}
\begin{document}


\title{Superconductivity mediated by nematic fluctuations -- the dispersion of collective modes}
\author{Kazi Ranjibul Islam}
\email{islam074@umn.edu}
\affiliation{Department of Physics, University of Wisconsin-Milwaukee, Milwaukee, Wisconsin 53201, USA}

\author{Andrey Chubukov*}
\email{achubuko@umn.edu}
\affiliation{School of Physics and Astronomy and William I. Fine Theoretical Physics Institute,
University of Minnesota, Minneapolis, MN 55455, USA}
\begin{abstract}
We  analyze the spectrum of collective modes  in a superconductor, in which pairing is mediated by long-range nematic fluctuations.  Previous experimental and theoretical studies have found that a superconducting gap in such a system 
 is highly anisotropic and at any finite $T < T_c$ vanishes on four arcs of the Fermi surface even when the pairing symmetry is $s-$wave ($s^{+-}$ between hole and electron pockets). We  
 derive the expression for the  pair susceptibility $\chi(\bq,\Omega)$ at finite momentum $\bq$ and frequency $\Omega$ deep in the superconducting phase.   We analyze the spectral function, $\text{Im}\chi(\bq,\Omega)$, and its pole structure in  transverse (phase)  and longitudinal (amplitude)  channels and compare the results with those of a conventional $s$-wave superconductor. We find that the analytic structure of the pair susceptibility 
 in both channels 
 is qualitatively distinct from that in a BCS superconductor.  This gives rise to  highly unconventional 
 dispersion of  phase and amplitude collective modes.
\end{abstract}
\maketitle
\tableofcontents 
\section{Introduction}
The interplay between superconductivity and electronic nematicity has emerged as one of the central themes in the study of a number of correlated quantum materials. Nematic order—characterized by spontaneous breaking of point-group rotational symmetry~\cite{fradkin2010nematic}—appears ubiquitously across several families of unconventional superconductors, including the 
iron-based superconductors \cite{chu2012divergent,bohmer2013nematic,gallais2013observation,yi2017role,chu2010plane,mirri2015origin,chuang2010nematic}, cuprates \cite{ando2002electrical,hinkov2008electronic,sato2017thermodynamic,lawler2010intra,daou2010broken,wu2017spontaneous} and twisted bilayer graphene \cite{cao2021nematicity}. Among these, iron-based chalcogenide material FeSe and its isovalently doped counterparts, FeSe$_{1-x}$S$_x$ and FeSe$_{1-x}$Te$_x$, have attracted significant interest 
 as in distinction to other Fe-pnictide materials,  they display nematicity and unconventional superconductivity without 
    long-range magnetic order. 
  A pure FeSe possesses
long-range nematic order below $T_p \sim 85K$  and superconductivity below $T_c \sim 8K$.   
       Upon doping by S or Te, $T_p$ decreases and vanishes
 at 
  some 
  critical $x_\text{c}$ (0.17 for S doping and 0.53 for Te doping). 
Moreover, numerous  experimental data  \citep{hanaguri2018two,shibauchi2020exotic,sato2018abrupt,ishida2022pure,mukasa2023enhanced,xu2016highly,liu2018orbital,sprau2017discovery,nagashima2022discovery,
walker2023electronic,nag2024highly,mizukami2021thermodynamics,mizukami2023unusual,matsuura2023two}
pointed out 
 that superconductivity at $ x \geq x_c$ is qualitatively different from that in pure and weakly doped FeSe ($x < x_c$).   The evidence includes, e.g.,  change of gap structure between $x <x_c$ and $x \geq x_c$
 ~ \citep{xu2016highly,liu2018orbital,sprau2017discovery,nagashima2022discovery,
walker2023electronic,nag2024highly}, variation of the magnetic field dependence of $T_c$ \citep{mukasa2023enhanced},   temperature variation of the specific heat at $T < T_c$  \citep{sato2018abrupt,mizukami2021thermodynamics,mizukami2023unusual}, large residual density of states 
\citep{mizukami2021thermodynamics,nag2024highly} and  small superfluid density~\citep{matsuura2023two} at $x \geq x_c$.
 These data fueled speculations that the pairing glue at $x <x_c$ and at  $x \geq x_c$ may be different.
  In pure/weakly doped FeSe,  antiferromagnetic order does not exist at ambient pressure but emerges under a small increase of pressure. In this situation,  the prevailing belief is that the pairing glue is still provided by  spin fluctuations, like in other Fe-based superconductors. 
  However,  at  $x \geq x_c$  long -range magnetism does not develop even at strong pressure, and spin fluctuations are likely too weak to be the pairing glue.
  
  It has been pointed out, both microscopically and phenomenologically \cite{islam2024unconventional,islam2025unconventional,lederer2015enhancement,lederer2017superconductivity,yamase2013superconductivity,agatsuma2016structure} that experimental results for doped FeSe at $x \geq  x_c$ can be understood  assuming that the pairing is mediated by nematic rather than spin fluctuations. 
      In the vicinity of a generic nematic critical point, nematic fluctuations are strong and peaked at momentum transfer
      ${\bf k}_F - {\bf p}_F \approx 0$, where ${\bf k}_F$ and ${\bf p}_F$ are Fermi surface momenta. 
       This gives rise to near-equal attraction in different pairing channels: $s-$wave, $p-$wave, $d-$wave 
       (Refs. \cite{lederer2015enhancement,klein2018superconductivity}.
    The peculiarity of doped FeSe is that in this system  hole and electron Fermi surfaces are made of different $d-$wave Fe-orbitals. As the   consequence,  the pairing interaction mediated nematic fluctuations also contains strong dependence on 
       $({\bf k}_F + {\bf p}_F)/2 \approx {\bf k}_F$.  The latter gives rise to strong angular variation  of the gap function along the Fermi surface even if the pairing symmetry is $s-$wave (ref. \cite{islam2024unconventional}).
       The anisotropy is most prominent on the hole pocket, which is made from $d_{xz}$ and $d_{yz}$ orbitals.
        In  a recent publication \cite{islamUnconventionalprb}
       the two us explored the consequences of  strong 
      angular variation of the superconducting  gap on various thermodynamic, spectroscopic and transport observables, e.g., specific heat, Raman response, and optical conductivity. 
      In the present work, we  extend our analysis to another direction- the structure of collective excitations in  
      a nematic-fluctuation-mediated superconductor (NFMS).
 
 Collective excitations in conventional superconductors are well understood;  $s-$wave, $p-$wave and  $d-$wave gap symmetries are the best-studied cases \cite{bardasis1961excitons,Anderson1958b,NNB1958, AndersonGauge, Volkov1975,SchmidSchon1975, ASchmid,VolkovKogan1973,Kulik1981,barlas2013amplitude,vollhardt1990superfluid,benfatto2001phase,anderson1958random,combescot2006collective,katsumi2018higgs,littlewood1982amplitude,maiti2013s+,maiti2015collective,paramekanti2000effective,phan2023following,podolsky2011visibility,althuser2025collective,schwarz2020classification,sharapov2001finite,sharapov2002effective,shimano2020higgs,udina2019theory,yang2020theory,islam2026spatially}. They 
  describe fluctuations of a 
  complex superconducting order parameter $\Delta(\br, t) = |\Delta|(\br,t) e^{i\phi(\br,t)}$. Transverse (phase) and longitudinal (amplitude) modes 
  describe fluctuations  of  $\phi(\br,t)$ and  $|\Delta|(\br,t)$, respectively.
  For a charge neutral s-wave superconductor with  a constant gap $\Delta$, the  phase mode is a gapless Goldstone mode \cite{bardasis1961excitons,AndersonGauge, Volkov1975,SchmidSchon1975, Kulik1981} with dispersion relation 
   $\Omega(\bq)= v_F |\bq|/\sqrt{2}$ (in $2d$) where $v_F$ is the Fermi velocity.
   It is  often called Anderson-Bogolubov (AB) mode \cite{Anderson1958b,NNB1958}. 
   The  longitudinal mode,  often called  Schmid-Higgs (SH) mode ~\cite{ASchmid,VolkovKogan1973,Galaiko1972,Galperin1981,Shumeiko1990,Spivak2004,Andreev2004},
is gapped. It  exists at  frequencies above $2\Delta$ and is a resonance in the continuum rather than a sharp mode.
    For a d-wave superconductors with the nodal gap structure the phase mode is not much different from the s-wave case,  but the  SH mode is significantly modified due to scattering off  gapless quasi-particles near the nodal points~\cite{paramekanti2000effective,podolsky2011visibility,sharapov2002effective,islam2026spatially}.

 In this communication, we analyze 
  collective excitations in a 2D  NFMS. We compute
 longitudinal $\chi_T (\bq, \Omega)$ and transverse $\chi_L (\bq, \Omega)$  particle-particle susceptibilities and 
  extract from them 
 the dispersion of the longitudinal and transverse collective modes. 
 For the convenience of a reader, we list here the key findings of our analysis. 
 
\begin{itemize}
\item In the transverse channel,  we find, neglecting Coulomb interaction, that there are no sharp  modes, but
still, $\mathrm{Im}\,\chi_T(\bq,\Omega)$  has two  maxima at $\Omega_1 = v_F |\bq| |\cos\theta_\bq|$ and $\Omega_2 = v_F |\bq| |\sin\theta_\bq|$, where $\theta_\bq$ specifies the  direction of $\bq$ with respect to $\hat x$. 
  The maxima merge at $\theta = \pi/4 (1+2n)$, but  even in this case the transverse mode remains damped.  
  We expect 
   that in the presence of the Coulomb interaction,
   the dispersion of the two maxima will change from $\Omega \propto q$ ro $\Omega \propto \sqrt{q}$, but
    the transverse modes remain 
    damped.  

 \item In the longitudinal channel, the  structure of collective excitations is different at $q=0$ and a finite $q$. At $q=0$, we find  that $\mathrm{Im}\,\chi_L(0,\Omega)$ is non-zero at all frequencies.  It has a 
 two-sided logarithmic singularity
 $ \mathrm{Im}\chi_L(0,\Omega)\propto \log(|\Omega-2\Delta_0|)$ at $\Omega = 2\Delta_0$, where $\Delta_0$ is the largest gap  on the Fermi surface.  At smaller $\Omega$, it remains weakly $\Omega$ dependent down to exponentially  small $\Omega$ and then drops as $1/(\log{(\Delta_0/\Omega)})^{1/2}$.  The real part of the longitudinal susceptibility 
 $ \mathrm{Re}\chi_L(0,\Omega)$ jumps by a finite value at $\Omega = 2\Delta_0$ and diverges at $\Omega \to 0$ as 
 $(\log{(\Delta_0/\Omega)})^{1/2}$. 
 
\item At a finite $\bq$,  the  logarithmic singularity
 in  $ \mathrm{Im}\chi_L(0,\Omega)$ at $\Omega =2\Delta_0$  splits 
 into two log-singularities at 
 \begin{align}
  \Omega_{\text{peak},1}(\bq)  &= 2\Delta_0 + \frac{v_F^2 |\bq|^2}{4\Delta_0} \cos^2\theta_{\bq},\quad 
  \Omega_{\text{peak},2}(\bq) = 2\Delta_0 + \frac{v_F^2 |\bq|^2}{4\Delta_0} \sin^2\theta_{\bq}.
  \end{align}
 The real part $ \mathrm{Re}\chi_L(0,\Omega)$  jumps  by a finite value  at these two frequencies. 
 
At small $\Omega$, $\mathrm{Im}\chi_L(q,\Omega)$ vanishes below an angle-dependent  threshold frequency, proportional to $q$.  For  $\theta_q =0$ and $\theta_q = \pi/4$, it   jumps to a finite value at this frequency and rapidly moves up around another characteristic frequency, also proportional to $q$.   For 
 $0<\theta_q < \pi/4$, there are two true jumps in $\mathrm{Im}\chi_L(q,\Omega)$ .
The real part $ \mathrm{Re}\chi_L(0,\Omega)$ has two maxima at small $\Omega$, where   
$\mathrm{Im}\chi_L(q,\Omega)$ either jumps or almost jumps. 
\end{itemize}

This behavior is qualitatively distinct from  
 that in both gapful and nodal conventional superconductors,  and  should be detectable  in spectroscopic probes such as Raman scattering or THz conductivity.

The paper is organized as follows. in Sec.~\ref{Model Hamiltonian section}, we present our model Hamiltonian and obtain the superconducting gap function at zero temperature. In Sec~\ref{Sec: Pairing susceptibility} we compute the particle-particle dynamic  susceptibility $\chi(\bq,\Omega)$ and split it into transverse and longitudinal parts. 
We discuss the  transverse part in Sec.\ref{Sec: Phase mode} and the  longitudinal part in Sec.\ref{Sec: Longitudinal mode}. 
In Sec. \ref{Sec:comparison} 
 we compare the collective modes in NFMS with those in gapful  and nodal BCS-type superconductors 
 ($s-$wave and $d-$wave).  We present our conclusions in  Sec.~\ref{Sec: Conclusion}. 

\section{Model and Superconducting Gap Structure}
\label{Model Hamiltonian section}
We consider an effective one band model of 2D electrons with momentum-dependent short-range interaction  in the Cooper channel. 
The corresponding Hamiltonian is 
\begin{align}
H=\sum_{\bk,\sigma} \xi(\bk) c^\dagger_{\bk,\sigma} c_{\bk,\sigma}+
     \sum_{\bk,\bp,\bq} V(\bk,\bp) c^\dagger_{\bk,\uparrow} c^\dagger_{-\bk ,\downarrow} c_{-\bp ,\downarrow}c_{\bp,\uparrow}
    \label{Hamiltonian},
\end{align}
where $c_{\bk,\sigma}$ is a fermion annihilation operator at momentum $\bk$ and spin $\sigma= \uparrow,\downarrow$. 
With FeSe in mind, we approximate the fermionic  dispersion $\xi_{\bk}$ by isotropic and parabolic $\xi_k =\bk^2/ 2m -\mu$,  with effective band mass $m$ and chemical potential $\mu$.  We  restrict the  pairing interaction to 
   fermions on the Fermi surface $ V(\bk,\bp)\equiv V(\theta_\bk,\theta_\bp)$, where $\theta_\bk (\theta_\bp)$ is the angle between the direction of  momentum $\bk_F (\bp_F)$ and $k_x$ axis. We will use $\Lambda$ for  the ultra-violet cutoff for the  theory. 

The pairing interaction mediated by nematic fluctuations is attractive and generally has the form
$V(\theta_\bk,\theta_\bp) = -f^2 (\theta_k, \theta_p) \chi_{nem} (\theta_k - \theta_p)$,  where $\chi_{mem}$ is the nematic susceptibility, and  $f(\theta_k, \theta_p)$ is the nematic  form factor.  The form of $f(\theta_k, \theta_p)$ 
depends on the microscopic model and the origin of the nematic order.  In many Fe-based superconductors, nematic   transition occurs only a few Kelvin above a magnetic transition into a stripe-magnetic phase  with momenta $(\pi, 0)$   or $(0,\pi)$ , and the two transition lines follow each other as functions of doping. In these systems nematicity is believed to be a vestigial state in which the symmetry between  $(\pi, 0)$  and $(0,\pi)$ 
 is broken,  but a magnetic long-range order is not yet set.
 For this case, $f(\theta_k, \theta_p)$  does not have  strong angle-dependence and can be reasonably well approximated by a constant.   As we said above, in doped FeSe magnetic and nematic transitions are well separated.  We follow earlier works~\cite{lederer2015enhancement,schattner2016ising,lederer2017superconductivity,klein2018superconductivity,islam2024unconventional,islam2025unconventional} and assume that nematicity in FeSe is a spontaneous Pomeranchuk order.
  In the microscopic consideration, this order emerges  because of  multiple $d-$wave orbitals, most notably $d_{xz}$ and $d_{yz}$, and the primary nematic order parameter  is the difference between the occupation of these two orbitals.  In this situation, the form factor is the combination of  the coherence factors from the transformation from orbital to band basis.  Microscopic calculation in~\cite{islam2024unconventional,islam2025unconventional} 
   found $f(\theta_\bk, \theta_\bp) = A \cos({\theta_\bk + \theta_\bp})$.  Near the onset of a nematic order 
   in  FeSe$_{1-x}$S$_x$(Te$_x$)  (coming from larger $x$), $\chi_{nem} (\theta_k - \theta_p)$ is strongly peaked at $\theta_k =\theta_p$ and for the purpose of superconductivity, can be approximated by $\delta (\theta_k - \theta_p)$.  Then $f(\theta_k, \theta_p) = A \cos{2\theta_k} $ and the pairing interaction can be expressed as  
  \begin{align}
    V(\bk,\bp)\equiv V(\theta_\bk,\theta_\bp)=- V_0 \cos^2 {2\theta_\bk} \delta(\theta_\bk-\theta_\bp)
    \label{Pairing Interaction}
\end{align}
where $V_0>0$. 
A similar form of the pairing interaction has been proposed for one-band models on phenomenological grounds~\citep{lederer2015enhancement,schattner2016ising,lederer2017superconductivity,klein2018superconductivity}.  
 This interaction is local ($\theta_k = \theta_p$)  and for this reason is the same in 
all  pairing channels that  form irreducible representations of the $D_{4h}$ group for the square lattice (one- dimensional $A_{1g}$,$B_{1g}$,$B_{2g}$ and two-dimensional $E$).  At some distance to a nematic transition, 
the equivalence between different channels is lost~\cite{lederer2015enhancement}. It is also lost at the transition if we keep the dynamical part of the interaction~\cite{klein2018superconductivity}. In both cases, $s-$wave superconductivity is favored.  Here we follow these works and assume that the pairing symmetry is $s-$wave. 
Other pairing symmetries, particularly  the chiral $p+ip$, may become favorable if one keeps the bare repulsive interaction between fermions in addition to the nematic one.  The analysis of collective modes has to be done separately for each pairing symmetry. 

\begin{figure}[]
 \centering
    \subfigure[]{\includegraphics[width= 0.36 \textwidth]{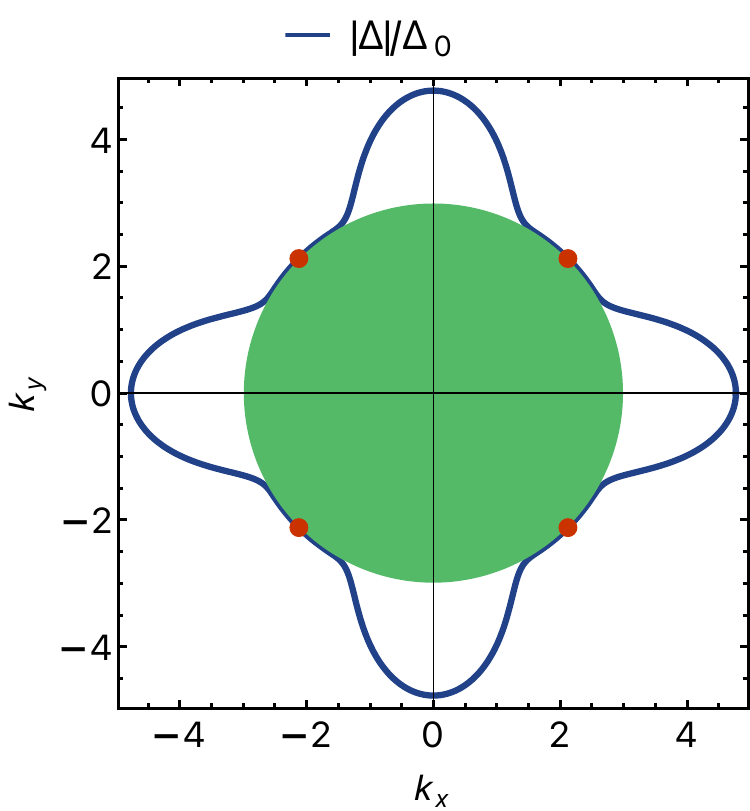}}
\caption{Polar plot for the gap function (blue curve) on the Fermi surface (green sphere) at zero temperature. In the cold region, the gap is non-zero except at points marked by red dots, but is exponentially small.}
\label{gap plot}
 \end{figure}

We now continue with $s-$wave pairing. 
The gap function $\Delta(\bk)$ at $T=0$ is obtained   by solving the non-linear gap equation 
\begin{align}
    \Delta(\theta_\bk)&= -\int \dfrac{d^2\bp}{(2\pi)^2} V(\theta_\bk,\theta_\bp) \,  \dfrac{ \Delta(\theta_\bp)}{2 \sqrt{\xi_\bp^2+|\Delta(\theta_\bp)|^2}},
    \label{Gap Equation 1}
\end{align}
For $V(\theta_\bk,\theta_\bp)$ given by \ref{Pairing Interaction}, 
this equation  becomes local
\begin{align}
    1= g \, \cos^22\theta_\bk \int_0^\Lambda  d \xi_\bk\dfrac{1}{ \sqrt{\xi_\bk^2+|\Delta(\theta_\bk)|^2}}
    \label{Gap at QCP}
\end{align}
where $g=N_0\, V_0$ is a dimensionless parameter, and $N_0=m/2\pi$ is the density of states per spin.
 Because the   strength of the attraction depends on $\theta_k$, the solution of \ref{Gap at QCP} is a highly anisotropic function of the direction of ${\bk}_F$
\begin{align}
    \Delta(\theta_\bk)=\Delta_0 \exp^{-\tan^22\theta_\bk/g}, \quad \Delta_0=2\Lambda \exp^{-1/g}. 
    \label{Gap form}
\end{align}
We plot the gap function as a function of angle on the Fermi surface in Fig.~\ref{gap plot}. It is peaked at angular positions $\theta_h= n \pi/2, n=0-3$, known as hot spots, and vanishes at $\theta_c=(2\,n+ 1) \pi/4$, known as cold spots. We stress that this holds even though the  gap symmetry is $s-$wave. 
Moreover, the gap  remains exponentially small in cold regions,  $\Delta(\theta\bk)\propto e^{-1/(4 g \delta\theta^2_{\bk})}$, where $\delta\theta_\bk=\theta-\theta_c$. The width of a cold region, where $\Delta \ll \Delta_0$,   
 is proportional to $\sqrt{g}$. 

\section{Pairing Susceptibility and Collective Modes}
\label{Sec: Pairing susceptibility}

We compute the dispersion of collective modes  using  Gorkov's diagrammatic technique.
 We do analytical calculations along the Matsubara axis and then rotate to  the real axis.
 
\begin{figure}[]
     \centering
     \includegraphics[scale=0.25]{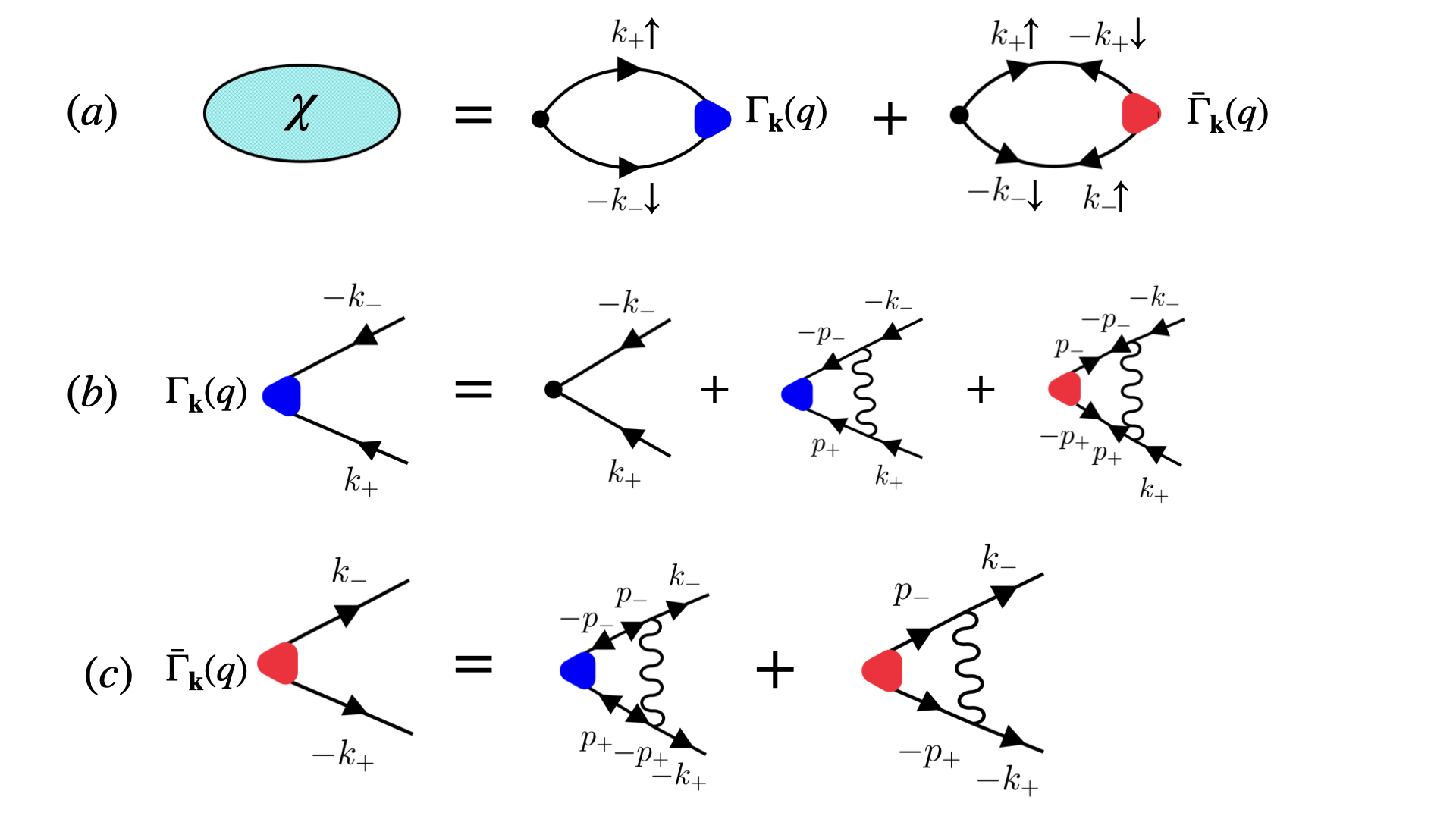}
     \caption{Dyson equation for (a) the pair-pair susceptibility $\chi(\bq,\Omega_m)$ with momentum $\bq$ and Matsubara frequency $\Omega_m$, (a) the renormalized particle-particle vertices $\Gamma_\bk$ (blue triangle with two incoming fermion momenta $k_+=k+q/2$ and $-k_-=-k+q/2$), (b) $\bar{\Gamma}_\bk$ (red triangle with two outgoing fermion momenta $k_-=k-q/2$ and $-k_+=-k-q/2$).   The black dot on the right vertex represents the s-wave form factor $1$, $\uparrow$ and $\downarrow$ label spin components of the internal fermion lines; the black single wavy line represents the pairing interaction. The single arrow solid lines represent normal Green's function $G$, and the double-headed arrow solid lines represent anomalous Green's function $F$. The intermediate momenta are labeled by $k_\pm=k\pm q/2$,  $p_\pm=p\pm q/2$ where $k=(\bk,\omega_k),p=(\bp,\omega_p)$ and $q=(\bq,\Omega_m)$.}
     \label{fig:vertex_figure}
 \end{figure}

 As it is customary in the Gorkov's technique, we  introduce normal ($G$) and anomalous ($F$) Green's functions as 
\begin{align}
 \label{Green's_function_exp}
G(\bk,\omega_m)=\dfrac{i\, \omega_m+\xi_\bk}{(i\, \omega_m)^2-E^2_{\bk}},\quad  F(\bk,\omega_m)=\dfrac{\Delta_\bk}{(i\, \omega_m)^2-E^2_{\bk}},
\end{align}
where $E_\bk=\sqrt{\xi^2_\bk+\Delta^2_\bk}$ is the quasi-particle excitation energy, $\Delta_\bk=\Delta(\theta_\bk)$ and 
 $\omega_m = \pi T (2m+1)$.   
 We compute diagrammatically
  the two-point pair-pair correlation function in the $s$-wave channel~\cite{Kulik1981}
 \begin{align}
     \chi
     (\bq,\Omega_m)&=\int_{-\infty}^\infty d\tau \, e^{i \Omega_m\tau} \int \dfrac{d^2\bk}{(2\pi)^2} \int \dfrac{d^2\bp}{(2\pi)^2}  \langle T_\tau \, c_{\bk+\bq/2,\uparrow}(\tau) c_{-\bk+ \bq/2,\downarrow}(\tau) c^\dagger_{-\bp+\bq/2,\downarrow}(0) c^\dagger_{\bp+\bq/2,\uparrow}(0) \rangle,
\label{chi_expression}
 \end{align}
where 
$\bq$ and $\Omega_m=2\pi mT$ . 
     We compute  $\chi (\bq,\Omega_m)$ \eqref{chi_expression} within the standard ladder/bubble approximation. Namely,  we sum up  series of ladder diagrams for the two-fermion vertices (see below) and use them to compute the dressed particle-particle bubbles. The corresponding diagrams are shown in Fig.\ref{fig:vertex_figure}.  The expression for $\chi
     (\bq,\Omega_m)$ in terms of single particle propagators 
      $G$ and $F$ is (see Fig.\ref{fig:vertex_figure}a )
\begin{align}
    \chi (\bq,i\Omega_m)&=\int_k  G(k+\dfrac{q}{2})\, G(-k+\dfrac{q}{2})\, \Gamma_q(\bk)- \int_k F(k+\dfrac{q}{2})\, F(-k+\dfrac{q}{2})\, \bar{\Gamma}_q(\bk),
    \label{chi_ecpression_2}
\end{align}
where the integration stands for $\int_k= N_0\,T\sum_{\omega_m} \, \int d\xi_\bk\int d\theta_\bk/2\pi$.
Below we compute the susceptibility at $T=0$,  in which case $ T\sum_{\omega_m} = \int d \omega_m/(2\pi)$. 

The equations for the two-particle 
vertices $\Gamma_q(\bk)$ and $\bar{\Gamma}_q(\bk)$ 
 are shown graphically in 
 Fig.~\ref{fig:vertex_figure} b,c. In analytical form 
 \begin{align}
    \Gamma_q(\bk)&=1-\int_k G(k+\dfrac{q}{2})G(-k+\dfrac{q}{2}) \Gamma_q(\bp)\, V(\bk,\bp)+ \int_k F(k+\dfrac{q}{2})F(-k+\dfrac{q}{2}) \bar{\Gamma}_q(\bp)\, V(\bk,\bp),\\
     \bar{\Gamma}_q(\bk)&=-\int_k G(k+\dfrac{q}{2})G(-k+\dfrac{q}{2}) \bar{\Gamma}_q(\bp)\, V(\bk,\bp)+ \int_k F(k+\dfrac{q}{2})F(-k+\dfrac{q}{2}) \Gamma_q(\bp)\, V(\bk,\bp).
\end{align}
Solving  these equations using the Green's functions from  $\eqref{Pairing Interaction}$, we  find that for $\bk$ on the Fermi ssurface 
\begin{align}
\label{Gamma_expression}
\Gamma_q(\bk)\equiv  \Gamma_q(\theta_\bk)&=\dfrac{1}{2}\left[\dfrac{1}{1-g \cos^22\theta_\bk \Pi_T(\theta_\bk,q)}+ \dfrac{1}{1-g \cos^22\theta_\bk \Pi_L(\theta_\bk,q)}\right], \\
\bar{\Gamma}_q(\bk)\equiv  \bar{\Gamma}_q(\theta_\bk)&=-\dfrac{1}{2}\left[\dfrac{1}{1-g \cos^22\theta_\bk \Pi_T(\theta_\bk,q)}- \dfrac{1}{1-g \cos^22\theta_\bk \Pi_L(\theta_\bk,q)}\right]. 
 \label{bar_Gamma_expression}
\end{align}
where  the angle-dependent $\Pi_{T,L}(\theta_\bk,q)$ are
\beq
\Pi_T(\theta_\bk,q)=\Pi_{GG}(\theta_\bk,q)+\Pi_{FF}(\theta_\bk,q), \quad \Pi_L(\theta_\bk,q)=\Pi_{GG}(\theta_\bk,q)-\Pi_{FF}(\theta_\bk,q)
\label{Pi_Transverse-expression}
\eeq
and $\Pi_{GG}(\theta_\bk,q)$ and $\Pi_{FF} (\theta_\bk,q)$ are  particle-particle bubbles made out of two G's and two F's, integrated over frequency and over $\xi$, but not over the direction of ${\bk}$.  We have
\begin{align}
\label{Pi_G}
    \Pi_{GG}(\theta_\bk,q)&= \int \dfrac{d\omega_\bk}{2\pi}\int d\xi_\bk  \, G(k+\dfrac{q}{2})\, G(-k+\dfrac{q}{2}),\\
   \Pi_{FF}(\theta_\bk,q)&= \int \dfrac{d\omega_\bk}{2\pi}\int d\xi_\bk   \, F(p+\dfrac{q}{2})\, F(-p+\dfrac{q}{2}).
   \label{Pi_F}
\end{align}
In explicit form,
\begin{align}
\Pi_{T/L}(\theta_\bk,q)=\dfrac{1}{4}\int_{-\Lambda}^{\Lambda} & d\xi_\bk \left(1+\dfrac{\xi_{\bk+\bq/2}\xi_{\bk-\bq/2}\pm \Delta^2(\theta_\bk)}{E_{\bk+\bq/2}E_{\bk-\bq/2}}\right) \times \nn &  \left[\dfrac{1}{E_{\bk+\bq/2}+E_{\bk-\bq/2}-i \,\Omega_m}+    \dfrac{1}{E_{\bk+\bq/2}+E_{\bk-\bq/2}+i\, \Omega_m} \right],
\label{Local_polarization_bubble}
\end{align}
where $E_\bk=\sqrt{\xi_\bk^2+|\Delta(\theta_\bk)|^2}$ is the quasi-particle excitation energy and $\Delta(\theta_\bk)$ is defined in Eq. ~\eqref{Gap form}. 
Substituting the expressions for $\Gamma_q(\theta_\bk),\bar{\Gamma}_q(\theta_\bk)$ ~\eqref{Gamma_expression}-\eqref{bar_Gamma_expression} into \eqref{chi_expression}, we obtain after some algebra the particle-particle susceptibility $\chi (\bq,\Omega_m)$  as the sum of two terms, which we identify as transverse ($\chi_T$) and longitudinal ($\chi_L$) susceptibilities: 
  \begin{align} 
   \chi (q) = \chi_T (q) + \chi_L (q),
   \end{align}
    where
\begin{align}
 \chi_T(q)=\dfrac{N_0}{2} \int \dfrac{d\theta_\bk}{2\pi}  \dfrac{\Pi_T(\theta_\bk,q)}{1-g \cos^22\theta_\bk \Pi_T(\theta_\bk,q)},\quad  
    \chi_L(q)=\dfrac{N_0}{2} \int \dfrac{d\theta_\bk}{2\pi}   \dfrac{\Pi_L(\theta_\bk,q)}{1-g \cos^22\theta_\bk \Pi_L(\theta_\bk,q)},
   \label{susceptibility into two modes}
\end{align}
We further simplify the expression for $\chi_{T/L}(q)$ in Eq.~\eqref{susceptibility into two modes} by incorporating the gap equation\eqref{Gap at QCP},  which is equivalent to $1=g \cos^22\theta_\bk \Pi_T(\theta_\bk,0)$. Introducing 
$ \delta\Pi_T(\theta_\bk,q)=\Pi_T(\theta_\bk,q)-\Pi_T(\theta_\bk,0), \, \delta\Pi_L(\theta_\bk,q)=\Pi_L(\theta_\bk,q)-\Pi_T(\theta_\bk,0)$, we obtain from (\ref{susceptibility into two modes})
\begin{align}
    \chi_{T,L}(q)=-\dfrac{N_0}{2\, g}\int_0^{2\pi}\dfrac{d\theta_\bk}{2\pi}\left[\dfrac{1}{g\,\cos^4 2\theta_\bk\, \delta\Pi_{T,L}(\theta_\bk,q)}+\dfrac{1}{\cos^22\theta_\bk}\right],
    \label{NFMS chi expression}
\end{align}

We compare this expression with that for a BCS  $s-$wave superconductor in Sec. \ref{Sec:comparison}. 

\subsection{Transverse Susceptibility }
\label{Sec: Phase mode}

We first compute the transverse pair-susceptibility $\chi_T(q)$. 
Converting Eq.~\eqref{Local_polarization_bubble}) to real frequencies,  we obtain $\delta \Pi_T(\theta_\bk,\bq,\Omega)$ 
in the form 
\begin{align}
   \delta \Pi_T(\theta_\bk,\bq,\Omega)&=-\dfrac{1}{4}\int_{-\infty}^\infty d\xi_\bk  \dfrac{E_{\bk+\bq/2}+ E_{\bk-\bq/2}}{E_{\bk+\bq/2}\, E_{\bk-\bq/2}} \dfrac{\Omega^2-\left(\xi_{\bk+\bq/2}-\xi_{\bk-\bq/2}\right)^2}{(\Omega+i\delta)^2-\left(E_{\bk+\bq/2}+E_{\bk-\bq/2}\right)^2}.
     \label{delta piT equation}
  \end{align}
 We direct  $\bq$ at an angle $\theta_\bq$ w.r.t to $\hat x$.  For small $\bq$, we expand $(\xi_{\bk+\bq/2}-\xi_{\bk-\bq/2})^2\approx v^2_F|\bq|^2 \cos^2(\theta_\bk-\theta_\bq)$ and $(E_{\bk+\bq/2} + E_{\bk-\bq/2})^2 \approx 4 \Delta^2(\theta_\bk)+4 \xi^2_\bk+v^2_F|\bq|^2 \cos^2(\theta_\bk-\theta_\bq)$  in the  integrand of Eq.~\eqref{delta piT equation}. Integrating over $\xi_\bk$ and expanding  in $|q|/k_F$ and $\Omega/\Delta(\theta_\bk)$ to  second order, we find $\delta\Pi_T (\theta_\bk,\bq,\Omega)=\left((\Omega+i\delta)^2- v^2_F|\bq|^2 \cos^2(\theta_\bk-\theta_\bq)\right)/4 \Delta^2(\theta_\bk)$. Substituting this expression into Eq.~\eqref{NFMS chi expression},  we find  
\begin{align}
    \chi_T(\bq,\Omega)=-\dfrac{2\, N_0}{g^2\, \Omega } \int \dfrac{ d\theta_\bk}{2\pi} \dfrac{\psi(\theta_\bq+\theta_\bk)}{\Omega+i\, \delta -v_F |\bq|\cos\theta_\bk},
    \label{simplified tranverse susceptibility}
\end{align}
where 
\beq
\psi(\theta_\bk)=\Delta(\theta_\bk)^2/\cos^42\theta_\bk
\eeq
A word of caution. The expansion  of $\delta\Pi_T(\theta_\bk)$ in $\Omega/\Delta(\theta_\bk)$ breaks down near a cold spot $\theta_k = \pi/4 (1 +2n)$ as the gap there is  exponentially small. In Appendix \ref{Appendix_B}, we derive the expression for $\delta\Pi_T(\theta_\bk)$ for such $\theta_k$ by expanding in $\Delta(\theta_\bk)/\Omega$. We verified 
 that the contribution to $\chi_T(\bq,\Omega)$ from this range of $\theta_k$ is  subleading  to the contribution from Eq. (\ref{simplified tranverse susceptibility}).  

We note that $\chi_T(\bq,\Omega)$ is $C_4$ symmetric with respect to rotations of $\theta_\bq$ by $m \pi/2$. Without loss of generality, we then restrict to the range $0 \leq \theta_\bq < \pi/4$. The 
 imaginary part of $\chi_T (\bq, \Omega)$ in this range is 
\begin{align}
    \text{Im}\chi_T(\bq,\Omega)&=\dfrac{N_0}{g^2\, \Omega} \int d\theta_\bk \psi(\theta_\bk+\theta_\bq) \delta(\Omega-v_F |\bq|\cos\theta_\bk)\nn & = \dfrac{N_0}{g^2}\dfrac{\Theta\left(1-\dfrac{|\Omega|}{v_F|\bq|}\right)}{\Omega \sqrt{v^2_F|\bq|^2-\Omega^2}}\left[\psi\left(\theta_\bq+\cos^{-1}\dfrac{\Omega}{v_F|\bq|}\right)+\psi\left(\theta_\bq-\cos^{-1}\dfrac{\Omega}{v_F|\bq|}\right) \right],
    \label{chi T expression}
\end{align}
\begin{figure}[]
    \centering
    \includegraphics[width=\linewidth]{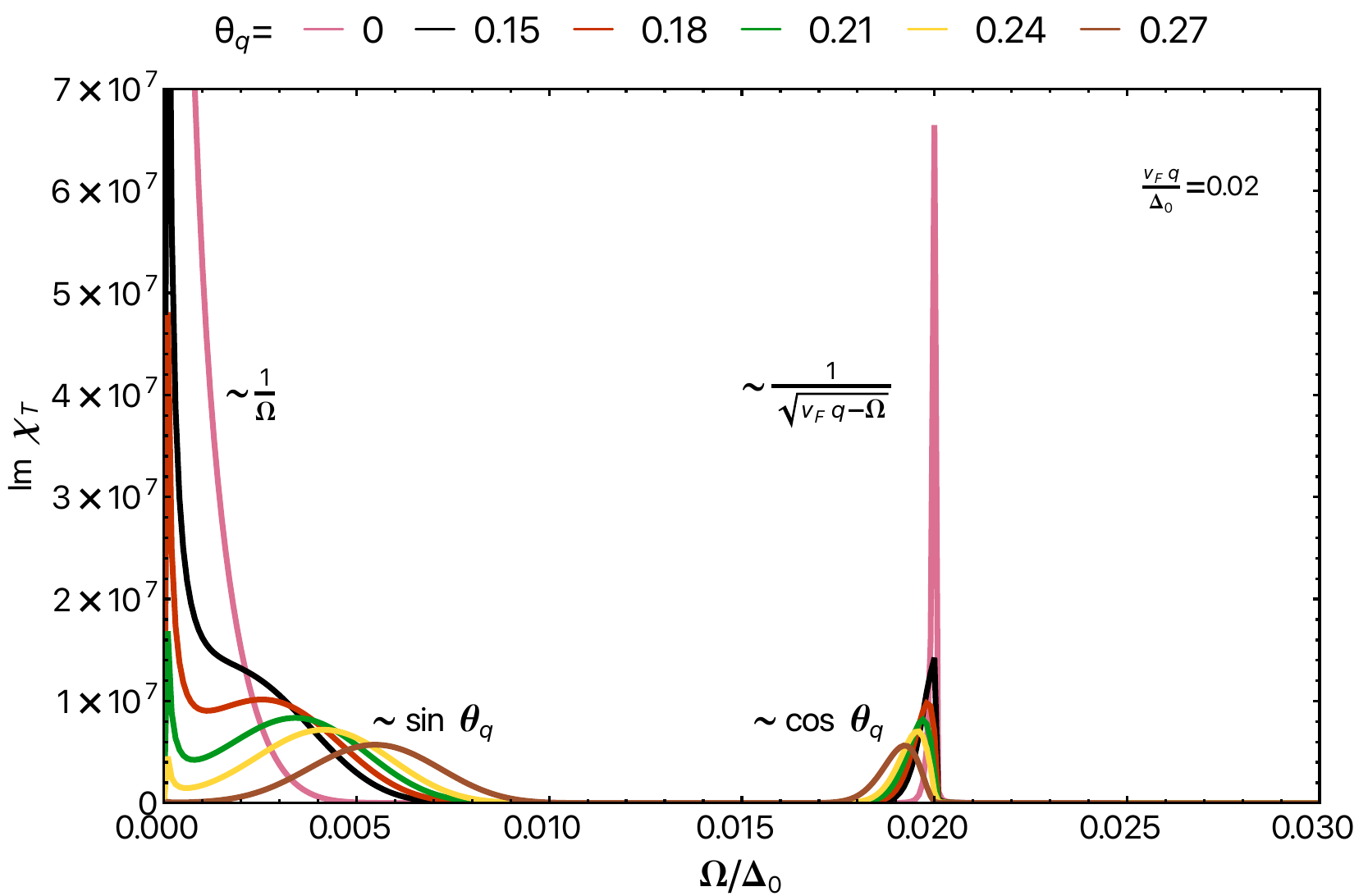}
    \caption{Imaginary part of the retarded transverse susceptibility $\chi_T(\bq,\Omega)$ as a function of frequency $\Omega$ for a fixed magnitude of momentum $\bq$. Different colors represent different directions of $\bq$, denoted as $\theta_\bq$. We mark special features of Im $\chi_T$ in different frequency regime by showing its $\Omega$ and $\theta_\bq$ dependence.}
    \label{fig: retarded susceptibility with thetaq}
\end{figure}
where  the Heaviside step function is defined as $\Theta (x) =0$ for $x<0$ and $\Theta(x) =1$ for $x >0$.
 In Fig.~\ref{fig: retarded susceptibility with thetaq}, we plot $\mathrm{Im}\,\chi_T(\bq,\Omega)$ as a function of frequency $\Omega$ for fixed $|\bq|$ and several values of $\theta_\bq$. 
 We see that it displays a rich structure.  First, $\mathrm{Im}\,\chi_T$ is nonzero for $\Omega < v_F |\bq|$. It diverges as $1/\Omega$ near the lower boundary and as $1/\sqrt{v_F |\bq| - \Omega}$ near the upper boundary. Second, because $\psi(\theta)$ is peaked at  $\theta=n\pi/2$ ($n=0,\dots,3$), $\mathrm{Im}\,\chi_T$ has two  peaks  at  $\Omega = v_F |\bq| \sin\theta_\bq$ and $\Omega = v_F |\bq| \cos\theta_\bq$. The two peaks have the same height $\Delta_0^2/(v_F^2 |\bq|^2 \sin 2\theta_\bq)$. The peak at $\Omega = v_F |\bq| \cos\theta_\bq$ is sharper, with a width proportional to $\sin\theta_\bq$, while the peak at $\Omega = v_F |\bq| \sin\theta_\bq$ is broader, with a width proportional to $\cos\theta_\bq$ (see Appendix.~\ref{Appendix_C} for details). At $\theta_\bq=\pi/4$, the two peaks merge at $\Omega = v_F |\bq|/\sqrt{2}$, 
  but even in this case the width of the peak is finite. 

\subsection{Longitudinal Susceptibility}
\label{Sec: Longitudinal mode}
\begin{figure}[]
  \centering
  \subfigure[]{\includegraphics[width = 0.47 \textwidth]{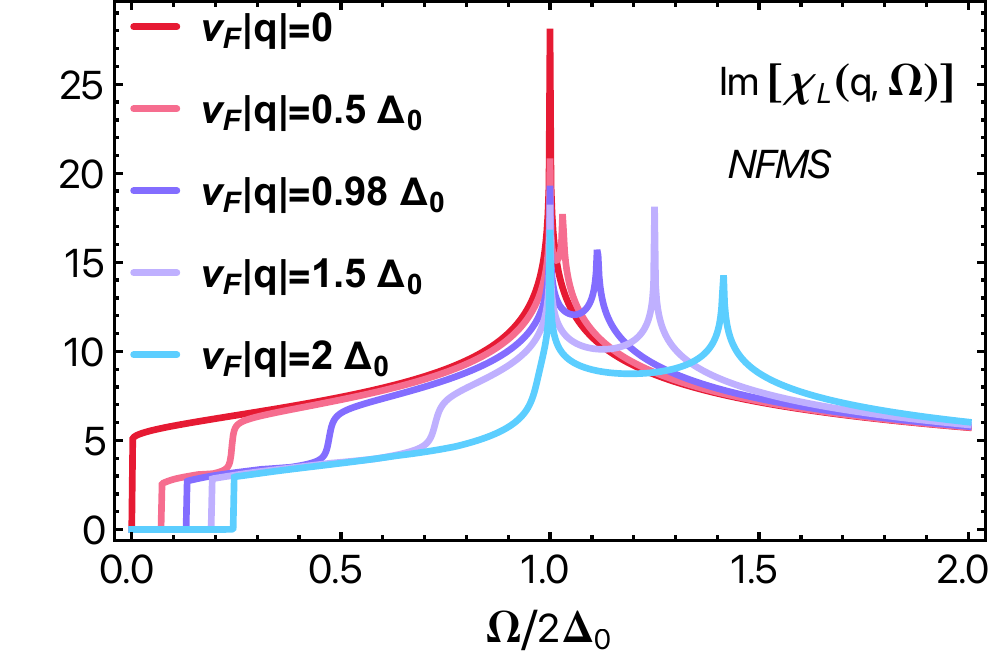}}
  \subfigure[]{\includegraphics[width = 0.47 \textwidth]{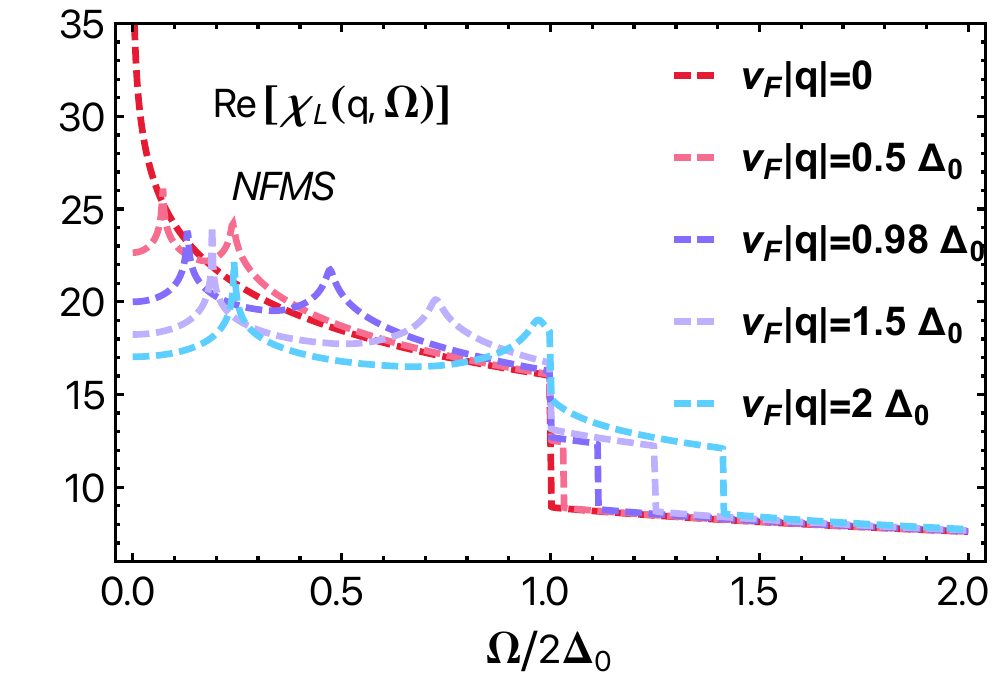}} 
 \caption{Frequency dependence of the imaginary (solid curve) and real parts (dashed curve) of the longitudinal susceptibility, evaluated at different values of momentum $\bq$ for 
 $\text{NFMS}$ for 
 $\bq$  alog ${\hat x}$.}
\label{Fig-chiSH}
\end{figure}

We next compute the longitudinal part of the  susceptibility,  $\chi_L$ from Eq.~\eqref{NFMS chi expression}. 
Using   Eq.~\eqref{Local_polarization_bubble}, we obtain the   
 retarded  polarization $\delta\Pi_L(\theta_\bk,\bq,\Omega)=\Pi_L(\theta_\bk,\bq,\Omega)-\Pi_T(\theta_\bk,0)$  
  in the form  
\begin{align}
    \delta \Pi_L(\theta_\bk,\bq,\Omega)&=-\dfrac{1}{4}\int_{-\infty}^\infty d\xi_\bk  \dfrac{E_{\bk+\bq/2}+ E_{\bk-\bq/2}}{E_{\bk+\bq/2}\, E_{\bk-\bq/2}} \,\dfrac{\Omega^2-4 \Delta^2(\theta_\bk)-\left(\xi_{\bk+\bq/2}-\xi_{\bk-\bq/2}\right)^2}{(\Omega+i\delta)^2-\left(E_{\bk+\bq/2}+E_{\bk-\bq/2}\right)^2}. 
     \label{delta_piL_equation_1}
\end{align}
 Approximating once again    $(\xi_{\bk+\bq/2}-\xi_{\bk-\bq/2})^2$ by $v^2_F|\bq|^2 \cos^2(\theta_\bk-\theta_\bq)$ and 
$(E_{\bk+\bq/2}+E_{\bk-\bq/2})^2$ by  $4 (\Delta_\bk)^2+4 \xi^2_\bk+v^2_F|\bq|^2 \cos^2(\theta_\bk-\theta_\bq)$ and   integrating over $\xi_\bk$, we obtain
\begin{align}
    \delta \Pi_L(\theta_\bk,\bq,\Omega)&=-\dfrac{\sec^{-1} \left(\dfrac{2\Delta(\theta_\bk)}{\sqrt{A^2(\theta_\bk,\bq)-(\Omega+i\delta)^2}}\right) \sqrt{A^2(\theta_\bk,\bq)-(\Omega+i\delta)^2}}{\sqrt{(\Omega+i\delta)^2-v^2_F |\bq|^2\cos^2\left(\theta_\bk-\theta_\bq\right)}},
     \label{delta piL equation}
\end{align} 
 where $A(\theta_\bk,\bq)=\sqrt{4 \Delta^2(\theta_\bk)+v^2_F|\bq|^2\cos^2(\theta_\bk-\theta_\bq)}$, and $\Delta(\theta_\bk)=\Delta_0 \exp\{-\tan^22\theta_\bk/g\}$. Inserting this expression into Eq.~\eqref{NFMS chi expression} and evaluating the remaining integrals numerically, we obtain the complex longitudinal susceptibility, $\chi_L(\bq,\Omega)$.  We show Re$\chi_L(\bq,\Omega)$ and Im$\chi_L(\bq,\Omega)$ in Fig.~\ref{Fig-chiSH} for $q$ along ${\hat x}$.  We see that both $\mathrm{Re}\,\chi_L(q,\Omega)$ and  $\mathrm{Im}\,\chi_L(q,\Omega)$ are non-zero for all frequencies. The structure of these two functions is different at $\bq =0$ and finite $\bq$.
 
 At $\bq=0$, $\mathrm{Im}\,\chi_L(\bq,\Omega)$  tends to a finite value as $\Omega$ decreases and  then lmost discontinuously 
  drops to  zero at $\Omega \to 0$ (the red curve in Fig.~\ref{Fig-chiSH}a). 
   Within the analytical treatment (see below),
    we find that the apparent discontinuity is actually the form
  $\mathrm{Im}\,\chi_L(\bq,\Omega) \propto 1/(\log{ (\Delta_0/\Omega)})^{1/2}$. It vanishes at $\Omega =0$ and rapidly increases at  exponentially small $\Omega$.  The  real part of $\chi_L(\bq,\Omega)$ diverges at $\Omega \to 0$  as
  $\mathrm{Re}\,\chi_L(\bq,\Omega) \propto (\log {(\Delta_0/\Omega)})^{1/2}$. The two forms are related by Kramers-Kronig transformation.  Near $\Omega =2\Delta_0$ $\mathrm{Im}\,\chi_L(\bq,\Omega)$ has a two-sided logarithmic singularity and  $\mathrm{Re}\,\chi_L(\bq,\Omega)$ jumps by a finite value. These two forms are also related by Kramers-Kronig.

At a finite $\bq$, the behavior is more complicated.  
At the smallest $\Omega$, 
    $\mathrm{Re}\,\chi_L(\bq,\Omega)$ tends to a finite value and $\mathrm{Im}\,\chi_L(\bq,\Omega)$ vanishes for  
      all $\theta_q$ except $\pi/4$. 
     The $\mathrm{Im}\,\chi_L(\bq,\Omega)$  jumps to a finite value at a critical, angle-dependent  $\Omega \propto q$, and $\mathrm{Re}\,\chi_L(\bq,\Omega)$ has a logarithmic singularity at this $\Omega$. 
     For $\theta_q =0$, shown  in   Fig.~\ref{Fig-chiSH},  there is a single jump  in $\mathrm{Im}\,\chi_L(\bq,\Omega)$  followed by a rapid increase  at a larger
       frequency, also proportional to $q$.    For  $\theta_q =\pi/4$, there is just a single jump.     For other $0<\theta_q <\pi/4$, there are two true jumps in  $\mathrm{Im}\,\chi_L(\bq,\Omega)$.  The real part of the longitudinal susceptibility  has a logarithmical singularity at the position of a true jump of $\mathrm{Im}\,\chi_L(\bq,\Omega)$  and a sharp maximum at a frequency around which $\mathrm{Im}\,\chi_L(\bq,\Omega)$  rapidly increases. 
   Near $\Omega = 2\Delta_0$,  $\mathrm{Im}\,\chi_L(\bq,\Omega)$  displays two logarithmic singularities at   
\begin{align}
\Omega_{\text{peak},1}(\bq)&=2\Delta_0+\frac{v_F^2|\bq|^2}{4\Delta_0}\cos^2\theta_\bq, \\
\Omega_{\text{peak},2}(\bq)&=2\Delta_0+\frac{v_F^2|\bq|^2}{4\Delta_0}\sin^2\theta_\bq.
\label{ex_aaa}
\end{align}
The real part  $\mathrm{Re}\,\chi_L(\bq,\Omega)$  undergoes two jumps  at these frequencies, as dictated by Kramers-Kronig relations.   

\begin{figure}[]
  \centering
 \subfigure[]{\includegraphics[width = 0.45 \textwidth]{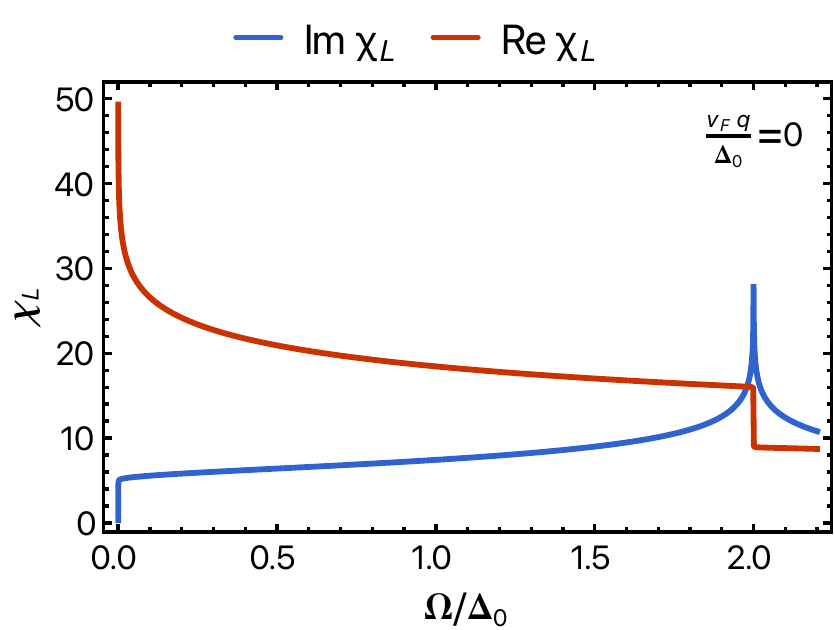}} \hspace{1 cm}
 \subfigure[]{\includegraphics[width = 0.47 \textwidth]{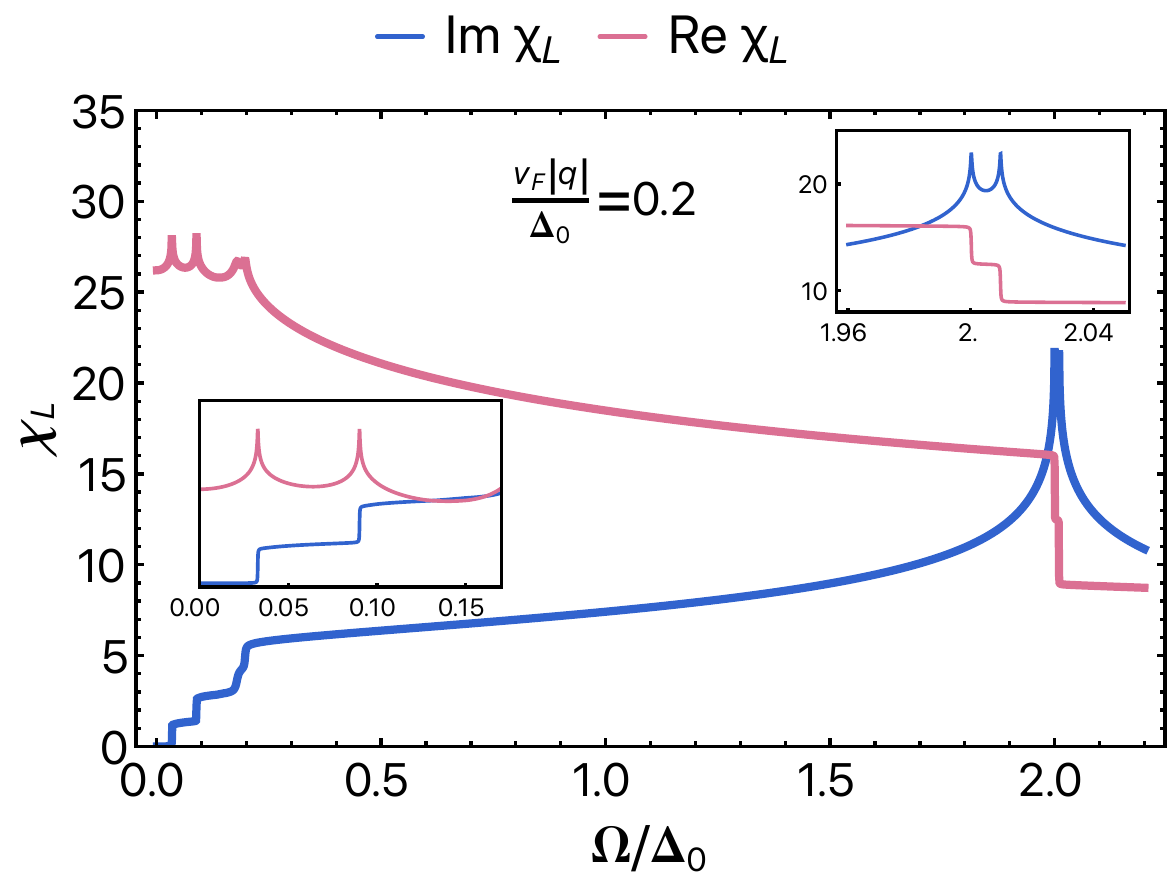}} 
 \caption{Frequency dependence of the real and  imaginary part of the retarded longitudinal pair-pair susceptibility $\chi_L(\bq,\Omega)$ for (a) $\bq$=0, and (b) $v_F \bq/\Delta_0=0.2$. In the insets of (b), behavior of $\chi_L(\bq,\Omega)$ is shown for small $\Omega$ and when $\Omega$ is near $2\Delta_0$, where $\Delta_0$ is the maximum gap amplitude. Here the direction of the momentum $\bq$ is chosen to be $\theta_\bq=\pi/20.$}
\label{fig:longitudinal susceptibility}
 \end{figure}

\begin{figure}[]
  \centering
 \subfigure[]{\includegraphics[width = 0.45 \textwidth]{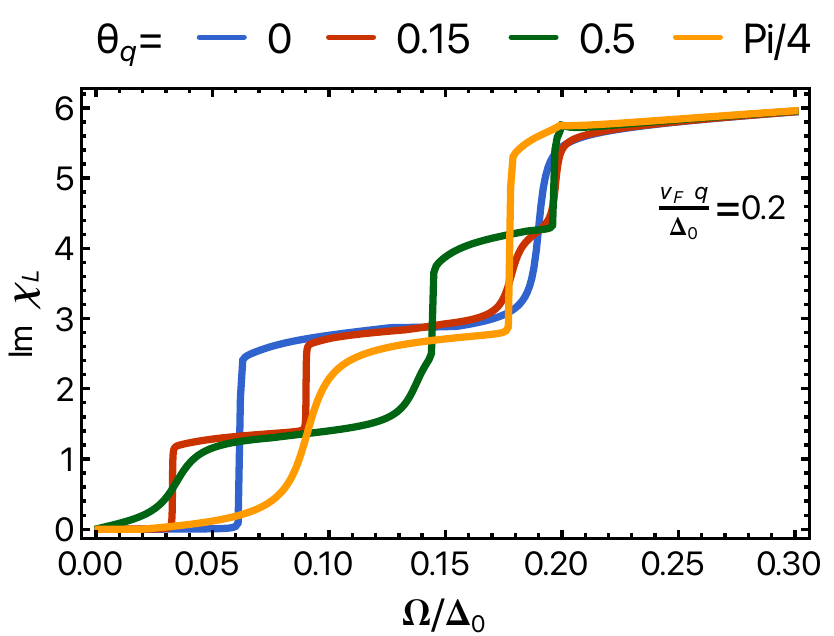}} \hspace{1 cm}
  \subfigure[]{\includegraphics[width = 0.45 \textwidth]{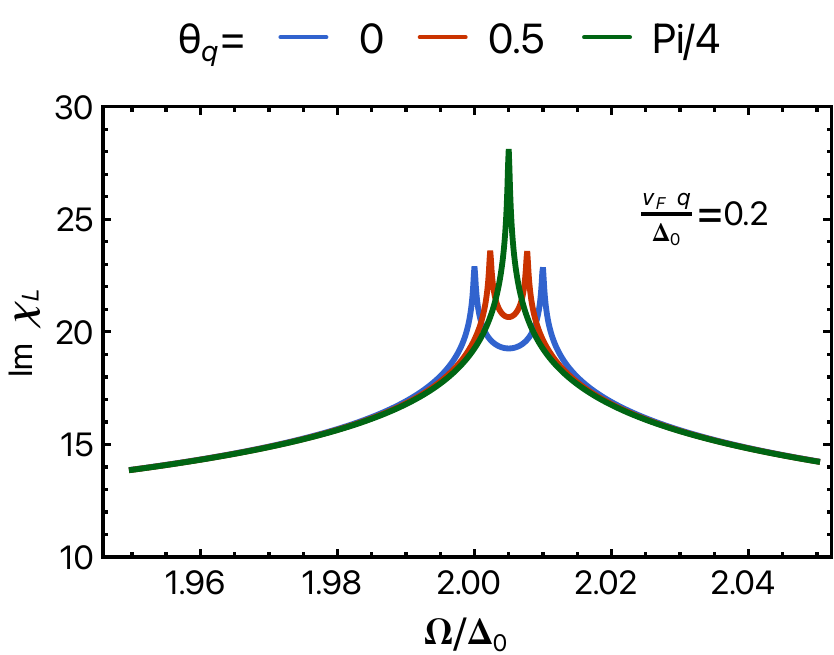}} 
  \subfigure[]{\includegraphics[width = 0.47 \textwidth]{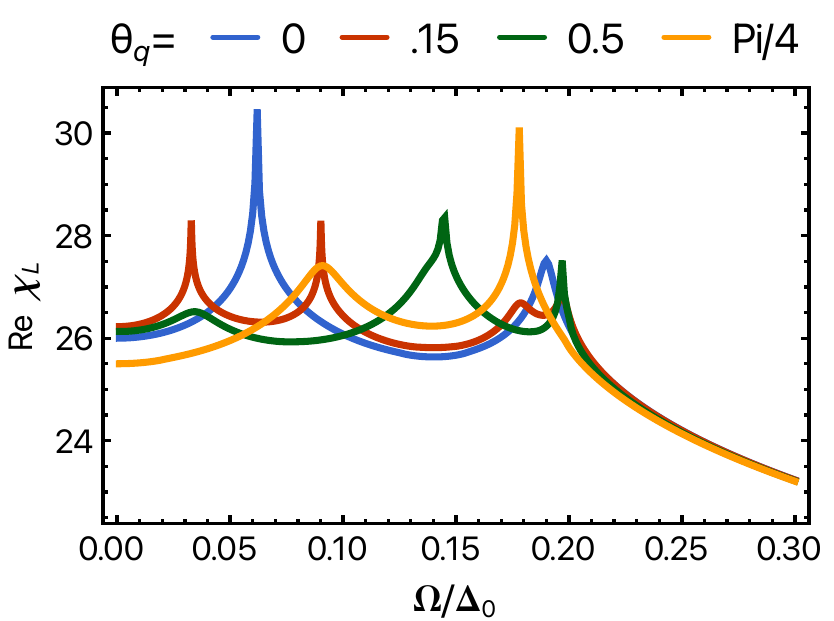}} 
  \subfigure[]{\includegraphics[width = 0.47 \textwidth]{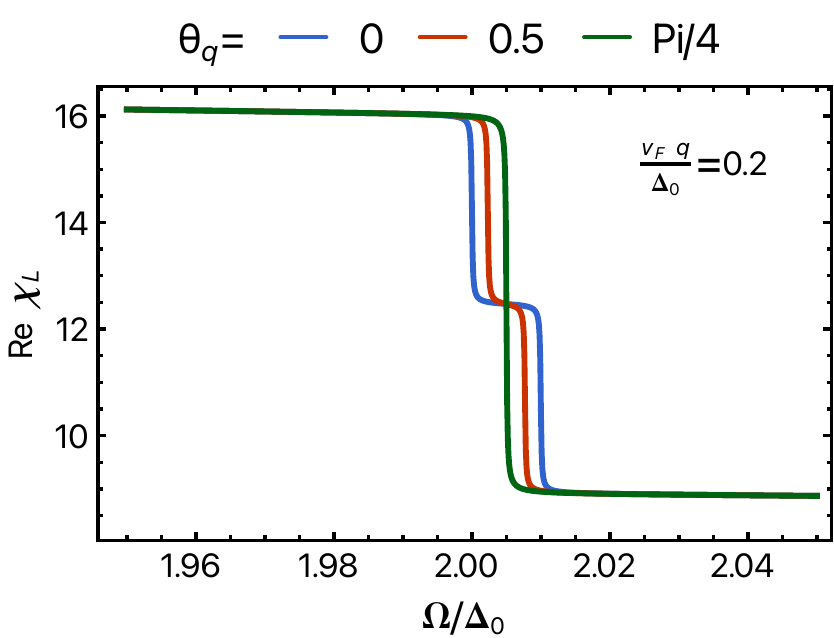}}
   \caption{Frequency dependence of the imaginary and real part of the retarded longitudinal pair-pair susceptibility, $\chi_L(\bq,\Omega)$ for (a,c) low frequency regime and (b,d) when frequency is close to $2\Delta_0$ for different values of momentum direction $\theta_\bq$. Here magnitude of $\bq$ is kept constant. At low frequency (left panel) $\text{Im}\chi_L(\bq,\Omega)$ shows discontinuities (or the  $\text{Re}\chi_L$ shows divergence) at in general two different frequencies (For $\theta_\bq=0, \text{and} \, \pi/4$ these frequencies are same, giving one discontinuity). Near $2\Delta_0$ (right panel) $\text{Im}\chi_L(\bq,\Omega)$ shows divenrgences (or the  $\text{Re}\chi_L$ shows discontinuities) at two different frequencies (For $\theta_\bq=\pi/4$ these frequencies are same, giving one divergence).}
\label{Fig_Anisotropic_chiL}
 \end{figure}

   We show  the behavior  at small $\Omega$ and at $\Omega \approx 2\Delta_0$ in more detail in Fig.  \ref{fig:longitudinal susceptibility}, where we compare 
 frequency dependencies of real and imaginary parts of $\chi_L(\bq,\Omega)$ at $q=0$ and at $v_F q =0.2 \Delta_0$, and 
 in Fig. \ref{Fig_Anisotropic_chiL},  where we compare the behavior of real and imaginary parts of $\chi_L(\bq,\Omega)$ 
 for different $\theta_q$.  We see that for $\theta_\bq =0$ and $\pi/4$, there is a single discontinuity in Im $\chi_l (\bq,\Omega)$ at small $\Omega$, while for other $\theta_q$ there are two discontinuities. We explain this result below in the section on the analytical analysis.   
 

\subsection{Analytical results}

Discontinuities and divergences of  $\text{Im}\chi_L(\bq,\Omega)$ displayed in 
 Figs. \ref{Fig-chiSH} - \ref{Fig_Anisotropic_chiL}
can be understood analytically  by analyzing Eq. \eqref{NFMS chi expression} and \eqref{delta piL equation}. We consider the cases $\bq =0$ and finite $\bq$ separately. 

 \paragraph{$\bq=0:$} 
At zero momentum, the longitudinal polarization bubble $\delta \Pi_L(\theta_k, 0, \Omega)$ still depends on $\theta_k$ due to angle-dependence of the gap function. We find
\begin{align}
\delta \Pi_L(\theta_\bk, 0, \Omega)
= - \dfrac{\sqrt{4 \Delta^2(\theta_\bk)-(\Omega+i\delta)^2}}{\Omega}
\sec^{-1}\left(\dfrac{2 \Delta(\theta_\bk)}{\sqrt{4 \Delta^2(\theta_\bk)-(\Omega+i\delta)^2}}\right),
\label{delta_piL_atq0}
\end{align}
where $\Delta(\theta_\bk)=\Delta_0 \exp\{-\tan^2(2\theta_\bk)/g\}$. At low frequencies, the dominant contribution to $\delta \Pi_L$ arises from the nodal regions where the gap is exponentially small. Expanding near a representative nodal point  $\theta=\pi/4$ as $\theta_\bk=\pi/4+\delta\theta_\bk$, and approximating $\Delta(\theta_\bk)\approx \Delta_0 \exp\{-1/(4g\,\delta\theta_\bk^2)\}$,  $\sqrt{4\Delta_\bk^2-\Omega^2}\approx -i \sqrt{2\Omega}\,\sqrt{\Omega-2\Delta(\theta_\bk)}$, we find 
\begin{align}
    \delta\Pi_L(\delta\theta_\bk,0,\Omega) \approx 
\log{\left(\dfrac{\Delta (\theta_\bk)}{\Omega}\right)} + i\dfrac{\pi}{2}  = -\dfrac{1}{4\, g\, \delta\theta^2_\bk} + \log\left(\dfrac{\Delta_0}{\Omega}\right) + i \dfrac{\pi}{2}
    \label{deltaPiL near cold spot}
\end{align}
substituting this expression into Eq. \eqref{NFMS chi expression} for $\chi_L(0,\Omega)$, approximating $\cos^2(2\theta_\bk)\approx 4\delta\theta_\bk^2$, and  evaluating  the integral over $\delta\theta_\bk$, we 
 obtain
\begin{align}
    \chi_L(0,\Omega)&\propto \int_0^{\delta\theta_0}  d\delta\theta_\bk \dfrac{1}{4 g \delta\theta^2_\bk} \left(\dfrac{1}{4 g \delta\theta^2_\bk \, \delta\Pi_L(\theta_\bk,0,\Omega)}+1\right)=\dfrac{1}{\sqrt{4g}} \left(\sqrt{\log \dfrac{\Delta_0}{\Omega}}+i \dfrac{\pi}{2 \sqrt{\log \dfrac{\Delta_0}{\Omega}}}\right).
    \label{chiL at small omega}
\end{align}
 This explains $(\log \dfrac{\Delta_0}{\Omega})^{1/2}$ behavior of Re $\chi_L(0,\Omega)$ and 
 $1/({\log \dfrac{\Delta_0}{\Omega}})^{1/2}$ behavior of Im $\chi_L(0,\Omega)$. 

 Near $\Omega = 2\Delta_0$, we introduce 
 $\Omega=2\Delta_0 (1 +\epsilon)$ and evaluate $\delta\Pi_L(\delta\theta_\bk,0,\Omega)$ at small $\epsilon$. We verified that for such $\Omega$,  the largest contribution to $\delta \Pi_L(\theta_\bk,0,\Omega)$ comes from the antinodal regions $\theta_{\mathrm h} \approx  n \pi/2, n=0-3$. Expanding the  gap function near an antinodal point 
 $\theta_\bk =0$ as   $\Delta_\bk=\Delta_0(1-\alpha \theta^2_\bk)$, where $\alpha=4/g$,
we obtain $\delta\Pi_L(\theta_\bk,0,\Omega)\equiv \delta\Pi_L(\theta_\bk,\epsilon)$ to the leading 
 order in $\epsilon$ and $\theta_k$ in the form
\begin{align}
    \delta\Pi_L(\theta_\bk,\epsilon)=\begin{cases}
        -\dfrac{\pi}{\sqrt{2}}\sqrt{|\epsilon|-\alpha\, \theta^2_\bk} & \epsilon<0 \\
        i\,  \dfrac{\pi}{\sqrt{2}}\sqrt{{\epsilon}+\alpha\,\theta^2_\bk} & \epsilon>0,
    \end{cases}
    \label{near 2delta0}
\end{align}
 Substituting this result into Eq.~\eqref{NFMS chi expression}, and approximating $\cos^2(2\theta_\bk)\approx 1$, we find that the imaginary part of $\chi_L(\epsilon)$  has a logarithmic singularity,

\begin{align}
     \text{Im}\chi_L(\epsilon) \propto \begin{cases}
         \dfrac{2\,\sqrt{2} N_0 }{\pi^2 \, g^2} \int_0^W d\delta\theta_\bk \dfrac{1}{\sqrt{ \alpha\, \delta\theta^2_\bk+\epsilon}} =\dfrac{\sqrt{2}N_0}{\pi^2 g^{2}\sqrt{\alpha}}\log \dfrac{4 W^2\alpha}{\epsilon}, & \epsilon>0  \\
         \dfrac{2\,\sqrt{2} N_0 }{\pi^2 \, g^2} \int_{\sqrt{|\Bar{\epsilon}|/\alpha}}^W d\delta\theta_\bk \dfrac{1}{\sqrt{ \alpha\, \delta\theta^2_\bk-|\Bar{\epsilon}|}} =\dfrac{\sqrt{2}N_0}{\pi^2 g^{2}\sqrt{\alpha}}\log \dfrac{32 W^2\alpha}{|\Bar{\epsilon}|}, & \epsilon<0,
     \end{cases}
     \label{imaginary part of chiL ar zero momentum}
\end{align} 
where  $W$ is the width of the hot region. 
The real part of $\chi_L$  has a discontinuity at $\Omega=2\Delta_0$:
\begin{align}
     \Delta\text{Re}\chi_L^R= \dfrac{2\,\sqrt{2}\, N_0}{\pi^2\, g^2}\int_0^{\sqrt{|{\epsilon}|/ \alpha}}  d\theta_\bk\dfrac{1}{\sqrt{|{\epsilon}|- \alpha \theta^2_\bk}}=\dfrac{N_0\, \sqrt{2}}{\pi g^2 \sqrt{\alpha}}.
     \label{real part of chiL ar zero momentum}
 \end{align}
Eqs.~\eqref{real part of chiL ar zero momentum}--\eqref{imaginary part of chiL ar zero momentum} satisfy Kramers--Kronig relations (see Appendix~\ref{Appendix_D} for details).\\

\paragraph{$\bq\neq 0:$}
We now analyze $\delta\Pi_L(\theta_\bk,\bq,\Omega)$, Eq. \eqref{delta piL equation},  at a finite $\bq$. 
We find analytically  that 
 $\mathrm{Im}\chi_{\mathrm{L}}(\bq,\Omega) =0$  below the threshold at 
 $\Omega_{min} = \textrm{min}\left\{A(\theta_\bk,\bq)\right\}$, where 
 \beq
 A(\theta_\bk,\bq)= \sqrt{4\Delta_\bk^2+v^2_F |\bq|^2 \cos^2(\theta_\bk-\theta_
     \bq)}.
     \eeq
     Analyzing  this condition, we find that $\Omega_{min}$ vanishes for 
    $\theta_q = \pi/4$, i.e., for $\bq$ along the nodal  direction, but is non-zero for all other values of $\theta_q$. The largest $\Omega_{min}$ is for $\bq$ along the antinodal directions ($\theta_q =0$, etc).
   
   We next analyze the origin of discontinuities and divergences in $\mathrm{Im}\,\chi_L(\bq,\Omega)$ at a finite $\bq$. These singular features arise when $\Omega$ matches  the local extrema of 
$A(\theta_\bk,\bq)$  w.r.t $\theta_\bk$.  Specifically, we show below that when $\Omega$ matches a local minima of $A$,
 $\mathrm{Im}\,\chi_L(\bq,\Omega)$ has a discontinuity, and when $\Omega$ matches a local maximum,  $\mathrm{Im}\,\chi_L(\bq,\Omega)$ diverges logarithmically. 
 By the Kramers--Kronig relations, these correspond respectively to divergence and discontinuity of $\mathrm{Re}\,\chi_L$. To quantify this behavior, we expand $\delta\Pi_L(\theta_\bk,\bq,\Omega)$ in Eq.~\eqref{delta piL equation} near an extremum of $A(\theta_\bk,\bq)$.  We express 
$A(\theta_\bk,\bq)=A_{\mathrm{ext}}(\bq)+\beta\,\delta\theta_\bk^2$, where
$\delta\theta_\bk=\theta_\bk-\theta_{\mathrm{ext}}(\bq)$,  $A_{\mathrm{ext}} (\bq) = A(\theta_{\mathrm{ext}}(\bq),\bq)$, and  $\beta >0$  at a minimum and $\beta <0$ at a maximum. 
Evaluating $\delta\Pi_L(\theta_\bk,\bq,\Omega)$, we obtain
\begin{align}
\delta\Pi_L(\delta \theta_\bk,\bq,\Omega)
=\dfrac{\pi}{2\sqrt{2}} \sqrt{A_{\mathrm{ext}}(\bq)}
\dfrac{\sqrt{-{\epsilon}+{\beta}\,\delta\theta_\bk^2}}
{\Delta(\theta_{\mathrm{ext}})},
\label{generic q case}
\end{align}
where $\epsilon=\Omega-A_{\mathrm{ext}}(\bq)$.  We first consider the case $\beta >0$ and set $\Omega$ to be near a 
 local minimum $A_{\mathrm{ext}}(\bq) \equiv A_\text{Min}(\bq)$. For $\epsilon<0$, $\delta\Pi_L$ is purely real, and hence $\chi_L$ remains real. For $\epsilon>0$, $\delta\Pi_L$ acquires an imaginary part within the range $\delta\theta_\bk < \sqrt{{\epsilon}/{\beta}}$. Substituting into Eq.~\eqref{delta piL equation}, we find that  
$\mathrm{Re}\,\chi_L(\bq,\Omega)$ jumps by a finite value at $\Omega_{\mathrm{disc}} = A_{\mathrm{Min}}(\bq)$:
\begin{align}
\Delta \mathrm{Im}\,\chi_L
= \dfrac{N_0 \sqrt{2}\, \Delta(\theta_{\mathrm{Min}})}{\pi^2 g^2 \sqrt{A_{\mathrm{Min}}(\bq)} \cos^4(2\theta_{\mathrm{Min}})}
\int_0^{\sqrt{{\epsilon}/{\beta}}}\,
\dfrac{ d\delta\theta_\bk}{\sqrt{{\epsilon}-{\beta}\,\delta\theta_\bk^2}}
= \dfrac{N_0\, \Delta(\theta_{\mathrm{Min}})}{\sqrt{2}\pi g^2 \cos^4(2\theta_{\mathrm{Min}})\sqrt{\beta\,A_{\mathrm{Min}}(\bq)}}.
\label{jump in imaginary part}
\end{align}
Consider next $\beta <0$ and set $\Omega$ to be close to  a local maximum $A_{\mathrm{ext}}(\bq) \equiv A_\text{Max}(\bq)$. Now  $\delta\Pi_L$ is purely imaginary for $\epsilon>0$. For $\epsilon<0$, it becomes real within $\delta\theta_\bk<\sqrt{|\epsilon/\beta|}$. This leads to a discontinuity in $\mathrm{Re}\,\chi_L$ at  $\Omega = A_\text{Max}(\bq)$. 
\begin{align}
\Delta \mathrm{Re}\,\chi_L
= \dfrac{N_0 \sqrt{2}\, \Delta(\theta_{\mathrm{Max}})}{\pi^2 g^2 \sqrt{A_{\mathrm{Max}}(\bq)}\cos^4(2\theta_{\mathrm{Max}})}
\int_0^{\sqrt{|{\epsilon}|/|{\beta}|}}\,
\dfrac{ d\delta\theta_\bk}{\sqrt{|{\epsilon}|-|{\beta}|\,\delta\theta_\bk^2}}
= \dfrac{N_0\, \Delta(\theta_{\mathrm{Max}})}{\sqrt{2}\pi g^2 \cos^4(2\theta_{\mathrm{Max}})\sqrt{\beta\,A_{\mathrm{Max}}(\bq)}},
\label{jump in real part}
\end{align}
By  Kramers--Kronig relation,  $\text{Im}\chi_L(\bq,\Omega)$ diverges logarithmically at 
 $\Omega=A_\text{Max}(\bq)$, which we label as 
 $\Omega_{\mathrm{peak}}$.

\begin{figure}[]
  \centering
  \subfigure[]{\includegraphics[width = 0.45 \textwidth]{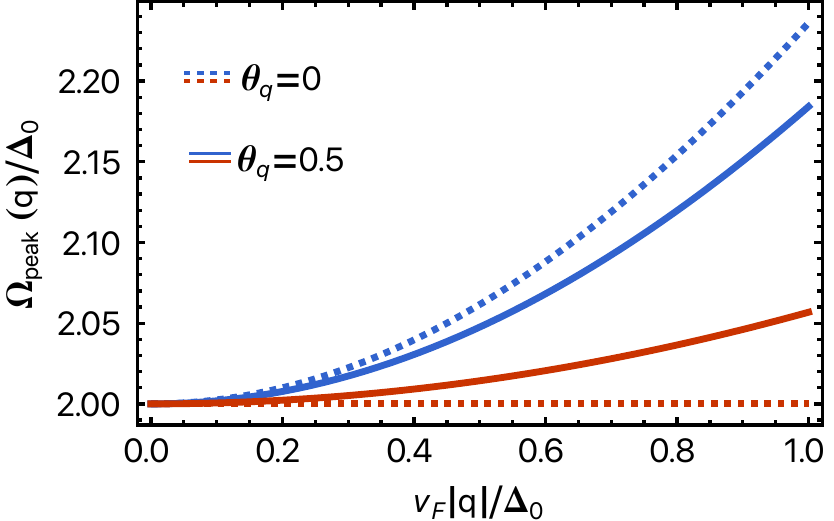}} \hspace{1 cm}
 \subfigure[]{\includegraphics[width = 0.47 \textwidth]{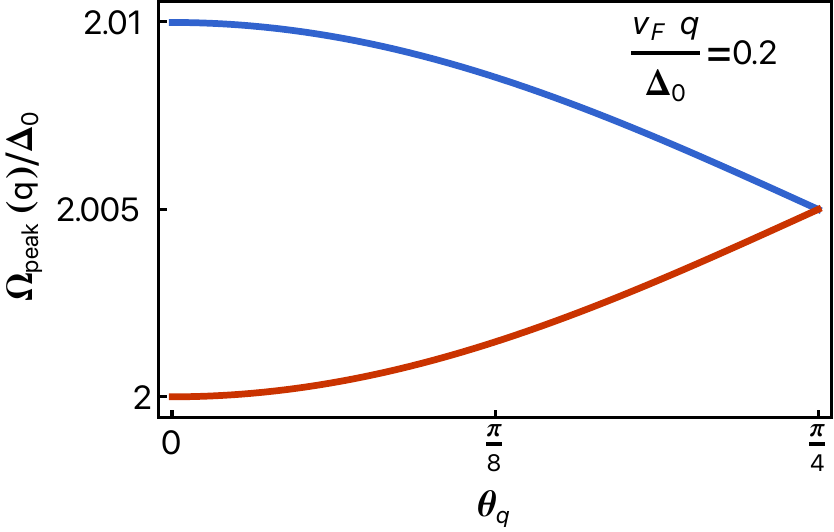}}
   \subfigure[]{\includegraphics[width = 0.47 \textwidth]{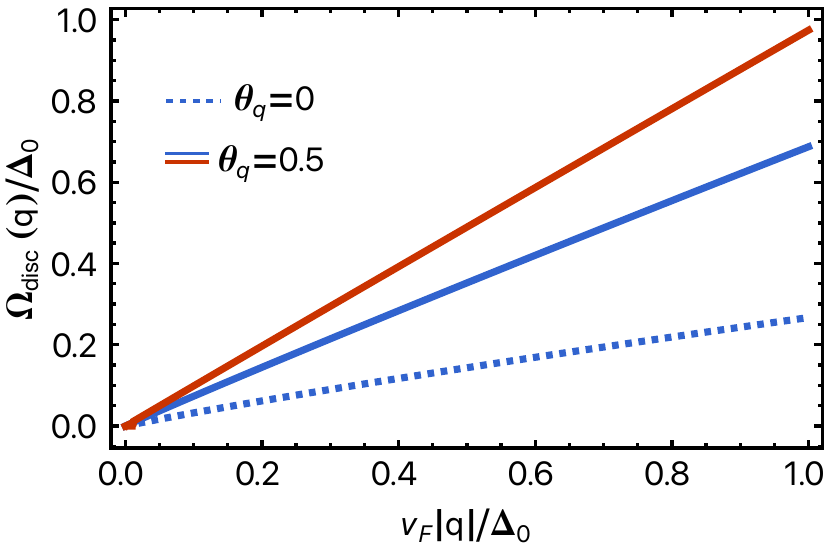}}
 \subfigure[]{\includegraphics[width = 0.45 \textwidth]{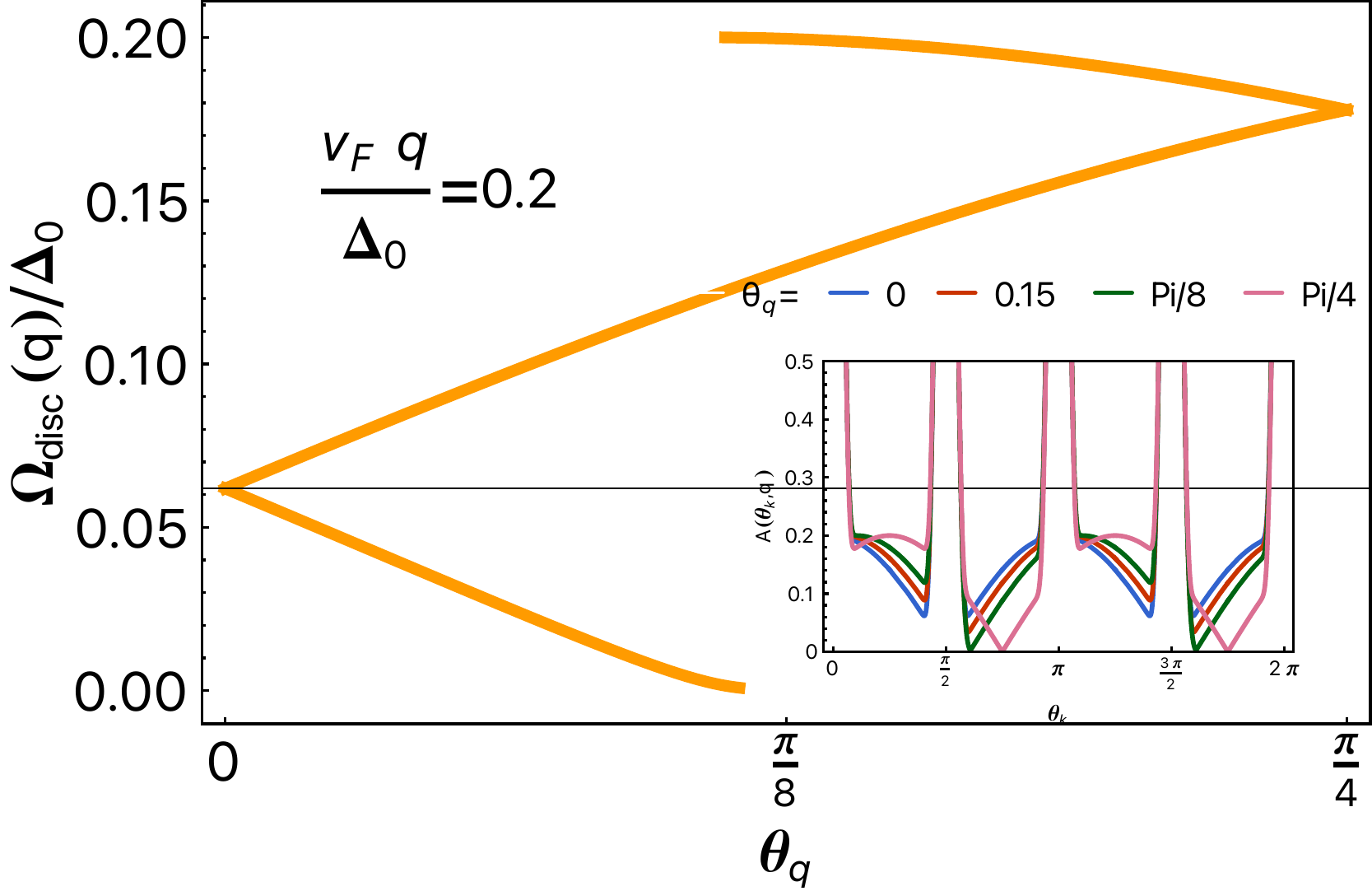}} 
  \caption{The momentum dependence of $\Omega_{\mathrm{peak}}(|\bq|,\theta_\bq)$ [panels (a,b)] and $\Omega_{\mathrm{disc}}(|\bq|,\theta_\bq)$ [panels (c,d)] is shown. The right panels display the dependence on $|\bq|$, while the left panels show the dependence on $\theta_\bq$. Here $\Omega_\text{peak}(\bq)$ is the frequency at which $\text{Im}\chi_L(\bq,\Omega)$ shows logarithmic divergent behavior, and $\Omega_\text{disc}(\bq)$ is the frequency at which $\text{Im}\chi_L(\bq,\Omega)$ shows discontinuous behavior. They are equal to the $A_\text{Max}(\bq)$ and $A_\text{Min}(\bq)$, respectively where $A_\text{Max/Min}$ is the extremum of the function $A(\theta_\bk,\bq)$ defined in the main text.  The inset in (d) shows the $\theta_\bk$ dependence of the function $A(\theta_\bk,\bq)$ for various directions of $\bq$.}
\label{chiL with q}
 \end{figure}

We now analyze the values of $\Omega_{\mathrm{disc}}$ and  $\Omega_{\mathrm{peak}}$ and how they depend on $\bq$. 
 In Fig.~\ref{chiL with q}, we plot  $\Omega_{\mathrm{peak}}(\bq)$ as a function of $|\bq|$ (panel (a)) and $\theta_\bq$ (panel (b)). For $\bq=0$, $A(\theta_\bk,0)=2\Delta(\theta_\bk)$ has maxima at the anti-nodal locations $\theta_{\mathrm{Max}}=n\pi/2$, where $\Delta (\theta_\bk) = \Delta_0$.  Expanding around this $\theta_\bk$, we find $\Omega_\text{peak}(0)=A_{\mathrm{Max}}(0)=2\Delta_0$ and  $\beta=2\Delta_0\,\alpha$, where $\alpha=4/g$. 
 The  discontinuity in  Re$\chi_L(0,\Omega)$ at $\Omega=2\Delta_0$ is $\sqrt{2}N_0/(\pi g^2 \sqrt{\alpha})$, in agreement with Eq.~\eqref{real part of chiL ar zero momentum}. For a small yet  finite $\bq$, $A(\theta_\bk,\bq)\approx  2 \Delta(\theta_\bk)+v^2_f|\bq|^2\cos^2(\theta_\bk-\theta_\bq)/4 \Delta_\bk$.
  The  local maxima of $A(\theta_\bk,\bq)$ are still  near the hot spots, but the value of  $A_\text{Max}(\bq)$ at $\theta_\text{Max}=0,\pi$ is different from that at $\theta_\text{Max} \approx \pi/2, 3\pi/2$:
   for $\theta_\text{Max} \approx 0,\pi$, $A_\text{Max}(\bq)=2\Delta_0+v^2_F|\bq|^2\cos^2\theta_\bq/4\Delta^2_0$, while for $\theta_\text{Max}=\pi/2, 3\pi/2$, $A_\text{Max}(\bq)=2\Delta_0+v^2_F|\bq|^2\sin^2\theta_\bq/4\Delta^2_0$. The  curvature remains  $\beta=2\Delta_0\alpha$ for both cases. As a result, $\mathrm{Im}\,\chi_L(\bq,\Omega)$ exhibits logarithmic divergence at two distinct frequencies $\Omega_{\text{peak},1}(\bq)$ and $\Omega_{\text{peak},2}(\bq)$,
   whose values we already presented in Eq. \ref{ex_aaa}.
 The jump in Re$\chi_L (\bq,\Omega)$ is same at $\Omega_{\text{peak},1}(\bq)$ and $\Omega_{\text{peak},2}(\bq)$
and equals $N_0/\sqrt{2}\pi g^2\sqrt{\alpha}$  - a  half of its value at $\bq=0$.
 When $\theta_\bq=n \pi/2, n=0-3$ (a direction towards a  hot spot), one of these frequencies remains
at $2\Delta_0$. When $\theta_\bq=(2n+1)\pi/4, n=0-3$ (a direction towards a cold spot),
 $\Omega_{\text{peak},1}(\bq) = \Omega_{\text{peak},2}(\bq)$.

We now analyze $A_{\mathrm{Min}}(\bq)$. For $\bq=0$, $A(\theta_\bk,0)=2\Delta(\theta_\bk)$ has no local minima, as the gap is exponentially flat near the cold spots and all angular derivatives vanish. Consequently, there is no discontinuity in $\mathrm{Im}\,\chi_L(0,\Omega)$. As we said, in this case 
 $\mathrm{Im}\,\chi_L(0,\Omega)$ is monotonic  and at the smallest $\Omega$ scales as  $1/\sqrt{\log(\Omega/\Delta_0)}$. At a small but finite $\bq$, $A(\theta_\bk,\bq)$ develops local minima. In Fig.~\ref{chiL with q}(c), we plot  $\Omega_{\mathrm{disc}}(\bq) = A_{\mathrm{Min}}(\bq)$. The minima disperse linearly with $|\bq|$ because they  originate from $\bk$  near cold spots, where $\Delta(\bk)$ is exponentially small and  $A(\theta_\bk,\bq) \approx  |\bq| |\cos(\theta_\bk-\theta_\bq)|$.  In Fig.~\ref{chiL with q}(d), we plot $\Omega_{\mathrm{disc}} =  A_{\mathrm{Min}}$ as a function of $\theta_\bq$ at fixed $|\bq|$. The inset shows $A(\theta_\bk,\theta_\bq)$ as a function of $\theta_\bk$ and  illustrates how the location of the minima evolves with $\theta_\bq$. For $\theta_\bq=0$, $A(\theta_\bk,\bq)$ exhibits a single minimum, leading to one discontinuity in $\mathrm{Im}\,\chi_L$ (Fig. \ref{Fig-chiSH} and the  blue curve in Fig.~\ref{Fig_Anisotropic_chiL}a).  For $\theta_\bq\neq 0$, two distinct minima appear, resulting in two discontinuities (red curve in Fig.~\ref{Fig_Anisotropic_chiL}a). As $\theta_\bq$ increases, one minimum vanishes near $\theta_\bq\sim \pi/8$, while another one emerges simulataneously at a different location maintaining two discontinuities. At $\theta_\bq=\pi/4$, the two minima again coincide and 
 Im$\chi_L$  displays a discontinuity only at a single frequency (orange curve in Fig.~\ref{Fig_Anisotropic_chiL}a).
 
\section{Comparison with  a  BCS  superconductor}
\label{Sec:comparison}

Collective modes in a conventional BCS $s$-wave superconductor have been extensively studied; see, e.g.,\cite{anderson1958random,anderson1958coherent,bardasis1961excitons,Anderson1958b,NNB1958, AndersonGauge, Volkov1975,SchmidSchon1975, ASchmid,VolkovKogan1973,Kulik1981, combescot2006collective,littlewood1982amplitude,VarmaLit1,maiti2013s+,maiti2015collective,phan2023following,podolsky2011visibility,althuser2025collective,schwarz2020classification,shimano2020higgs,udina2019theory}. Collective modes in a $d$-wave BCS superconductor have likewise been discussed in, e.g., \cite{barlas2013amplitude,benfatto2001phase,paramekanti2000effective, benfatto2001phase,katsumi2018higgs,sharapov2001finite,sharapov2002effective, yang2020theory,islam2026spatially}. Because the pairing symmetry in our case is $s$-wave, we primarily compare our results with those for a conventional $s$-wave BCS superconductor.  
For  convenience of a reader,  in  Appendix~\ref{Appendix_A} we  derive the expressions for transverse and longitudinal susceptibilities for an ordinary s-wave BCS superconductor  with  a  momentum-independent pairing interaction $V_s(\bk,\bp) = -V_0$ and a  constant gap 
 $\Delta(\theta_{\bk}) = \Delta_0$.   The expressions for  transverse and longitudinal components of the particle-particle  susceptibility, $\chi^s = \chi^s_T + \chi^s_L$ are 
\begin{align}
    \chi^s_{T,L}(q) = -\frac{N_0}{2 g}\left[\frac{1}{g \int_0^{2\pi}\frac{d\theta_{\bk}}{2\pi}\, \delta\Pi^s_{T,L}(\theta_{\bk},q)} + 1\right],
    \label{BCS susceptibility into two modes}
\end{align}
where $g = N_0 V_0$ and
$\delta\Pi^s_{T,L}(\theta_{\bk},q) = \Pi^s_{T,L}(\theta_{\bk},q) - \Pi_T(\theta_{\bk},0)$ has the same structure as $\delta\Pi_{T,L}(\theta_{\bk},q)$ in Eq.~(\ref{Local_polarization_bubble}), but with $\Delta(\theta_{\bk})$ replaced by $\Delta_0$.

A useful way to contrast this result with the corresponding expression for the nematic fluctuation--mediated superconducting (NFMS) state, Eq.~\eqref{NFMS chi expression}, is through an analogy with electrical networks. If each momentum point on the Fermi surface (labeled by $\theta_{\bk}$) contributes a local susceptibility $\chi(\theta_{\bk},q) \sim 1/\delta\Pi(\theta_{\bk},q)$, then the total susceptibility combines differently in the two cases. In the NFMS case, Eq.~\eqref{NFMS chi expression}, the contributions add directly, $\chi(q) \sim \int d\theta_{\bk}\, \chi(\theta_{\bk},q)$, analogous to resistors in series. In contrast, in the conventional BCS case, Eq.~\eqref{BCS susceptibility into two modes}, the inverse susceptibility adds, $1/\chi(q) \sim \int d\theta_{\bk}\, 1/\chi(\theta_{\bk},q)$, analogous to resistors in parallel.
 \begin{figure}[]
  \centering
  \subfigure[]{\includegraphics[width = 0.47 \textwidth]{NFMS_ImChiWq.pdf}}
  \subfigure[]{\includegraphics[width = 0.47 \textwidth]{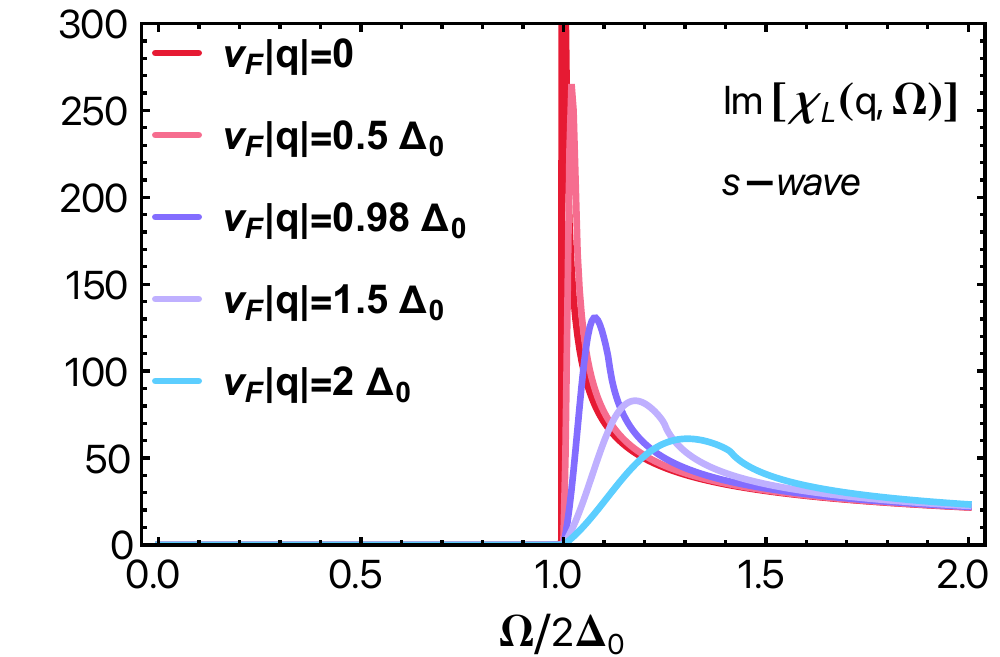}}
  \subfigure[]{\includegraphics[width = 0.47 \textwidth]{NFMS_ReChiWq.pdf}} 
 \subfigure[]{\includegraphics[width = 0.47 \textwidth]{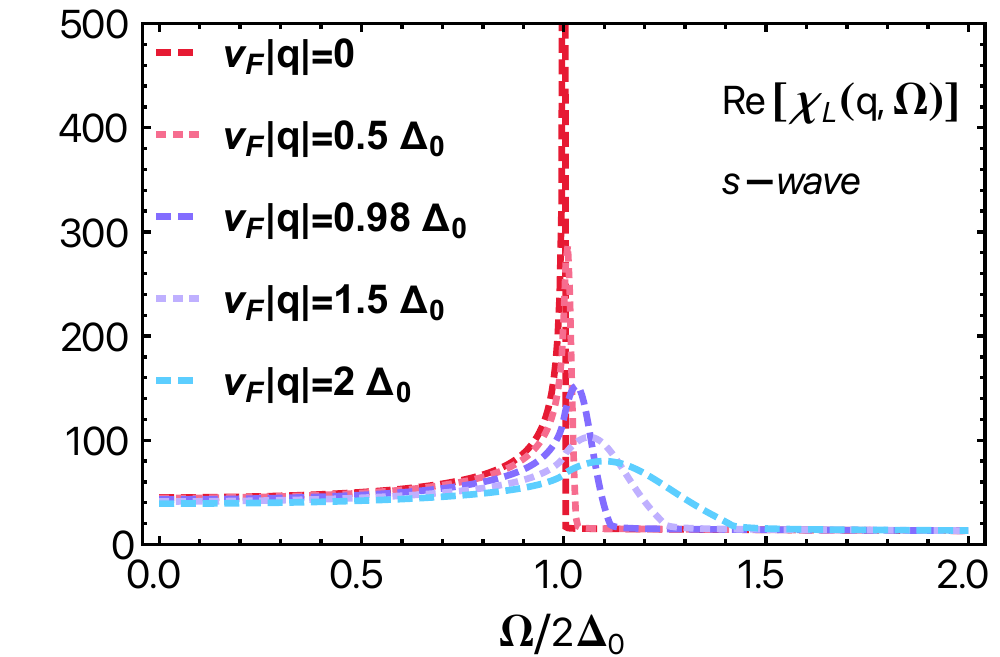}}
 \caption{Frequency dependence of the imaginary (solid curve) and real parts (dashed curve) of the longitudinal susceptibility, evaluated at different values of momentum $\bf q$  for a 
 $\text{NFMS}$ (right panel) and for  an $s$-wave BCS superconductor(left panel).
 The results for the $\text{NFMS}$  are for
 $\bq$  alog ${\hat x}$.}
\label{Fig-chiSH_1}
\end{figure}

In practical terms,  distinct structures of Eqs.~\eqref{NFMS chi expression} and \eqref{BCS susceptibility into two modes} lead to qualitatively different collective-mode spectra in the NFMS and  BCS $s$-wave cases.    In the transverse channel 
 the distinction is between a  sharp phase mode in a BCS $s-$wave superconductor with dispersion $\Omega = v_F |\bq|/\sqrt{2}$  ( $\Omega  \sim   |\bq|^{1/2} $ ) without (with) the Coulomb interaction 
  and two damped modes in NFMS at frequencies $\Omega_1 = v_F |\bq| |\cos\theta_{\bq}|$ and $\Omega_2 = v_F |\bq| |\sin\theta_{\bq}|$ without Coulomb interaction.
  In the longitudinal channel,  $\mathrm{Im}\,\chi^s_L (q,\Omega)$  in a BCS $s-$wave superconductor  vanishes below $2\Delta_0$ and 
  a square-root  singularity  at $\Omega = 2\Delta_0$:  $\mathrm{Im}\,\chi^s_L (0,\Omega) \propto \frac{1}{\sqrt{\Omega - 2\Delta_0}}$.   At a  finite $q$, $\mathrm{Im}\,\chi^s_L (q,\Omega)$   develops a peak at $\Omega > 2\Delta_0$,  which disperses quadratically with momentum \cite{Kulik1981,anderson1958random,anderson1958coherent,Anderson1958b,NNB1958,AndersonGauge,combescot2006collective,phan2023following,islam2026spatially}.   
  This is very  different from  $\mathrm{Im}\,\chi_L (q,\Omega)$  in  NFMS, which (i) remains non-zero below $2\Delta_0$, with  discontinuities at $\Omega \sim q$, and  (ii)   has logarithmic singularities  at $\Omega \approx 2\Delta_0$, which disperse with $q$ for a generic $\theta_q$ and do not disperse for special $\theta_q$.  We compare the behavior of the longitudinal susceptibility in NFMS and in a BCS $s-$wave superconductor in  
   Fig.~\ref{Fig-chiSH_1}. 
   
 The behavior of $\chi_L$  and $\chi_T$ in a NFMS is also qualitatively different from that  in a  d-wave BCS-type superconductor with 
  $\Delta(\theta_{\bk}) = \Delta_0 \cos 2\theta_{\bk}$.    There,  $\mathrm{Im} \,\chi^d_L (q,\Omega)  \sim \Omega^3$ at low frequencies due to point nodes~\cite{islam2026spatially,paramekanti2000effective,katsumi2018higgs,sharapov2001finite,sharapov2002effective, yang2020theory}. In NFMS the  frequency evolution  of $\mathrm{Im}\,\chi_L (0,\Omega)$  at small $\Omega$ 
    is much stronger due to exponentially small gap in the cold regions.   The behavior near $\Omega=2\Delta_0$ is also strikingly different. For the $d-$wave superconductor $\textrm{Im}\chi^d_{\mathrm{L}}(q,\Omega)$   is not-analytic but not singular, while in NFMS,  $\textrm{Im}\chi_{\mathrm{L}}(q,\Omega)$  has 
     two logarithmic singularities.

\section{Conclusions}
\label{Sec: Conclusion}

In this work, we investigate the collective mode dynamics in a superconductor in which the pairing is driven by soft charge-nematic fluctuations. Such a pairing scenario is especially relevant to FeSe$_{1−x}$S(Te)$_x$ systems, which host a nematic quantum critical point near $x_c\sim 0.17$ (for S-doping) and $x_c\sim 0.5$ (for Te-doping). Near $x_x$, strong nematic fluctuations, peaked at zero momentum, generate a highly anisotropic attractive  pairing interaction between electrons in the form $V(\bk,-\bk;\bp,-\bp)\propto -\cos^22\theta_\bk\, \delta(\theta_\bk-\theta\bp)$, where $\theta_\bk$ and $\theta_\bp$ are the angular position of the incoming/outgoing
fermions on the Fermi surface. The factor $\cos^22\theta_\bk$ is the effect of  coherent factors of the transformation from orbital to band basis (quantum geometry, in modern language). 
  This attraction gives rise to superconductivity with a  highly anisotropic  gap function $\Delta (\theta_k)$, even in the $s-$wave channel, which we consider here: $\Delta(\theta_\bk) = \Delta_0  \exp(-\tan^22\theta_\bk/g)$. The superconducting gap is maximized at the hot spots, $\theta_\bk= n \pi/2, n=0,1,2,3$,and is significantly suppressed in the cold regions  around $\theta_\bk= (2n+1) \pi/4$. We  analyze transverse and longitudinal  collective excitations of such a superconductor. For this, we  compute the pair-pair susceptibility $\chi(\bq,\Omega)$ and  decompose it into transverse ($\chi_T$) and longitudinal ($\chi_L$) components.  We find that the structure of the pair-pair susceptibility in our case is qualitatively different from that in conventional $s-$wave BCS superconductor, reminiscent of the difference between series and parallel resistor networks. 

  In the transverse channel, we found, neglecting the long-range Coulomb interaction,that 
$\mathrm{Im}\,\chi_T (\bq, \Omega)$ is non-zero in a finite range $0< \Omega< v_F |\bq|$. Within this range, it  has two  maxima  at $\Omega = v_F |\bq| \cos\theta_{\bq}$ and $\Omega = v_F |\bq| \sin\theta_{\bq}$, where $\theta_{\bq}$ defines the direction of the momentum $\bq$ relative to $\hat x$ towards a  hot spot. These peaks correspond to anisotropic sound-like damped collective modes, whose velocities scale linearly  with $\theta_{\bq}$.  We expect that in the presence of Coulomb interaction,  two damped modes remain, but with dispersion $\Omega \propto \sqrt{q}$.   

In the longitudinal channel, we found qualitatively different behavior at $q =0$ and at a finite $q$.
At zero momentum, $\mathrm{Im}\,\chi_L(0, \Omega)$ is nonzero  for all $\Omega$. It vanishes at $\Omega =0$, but  very rapidly increases at a non-zero $\Omega$, as $1/(\log(\Omega/\Delta_0)^{1/2}$.  We related this behavior 
 to the unique nodal structure of our $\Delta (\theta_k)$. Near  $\Omega = 2\Delta_0$, we found that 
 $\mathrm{Im}\,\chi_L (0, \Omega)$ diverges as $\log(|\Omega - 2\Delta_0|)$. This gives rise to a discontinuity in the real part of $\chi_L$ at the same frequency.

At a finite $q$, we found that at small $\Omega \ll \Delta_0$, $\mathrm{Im}\,\chi_L(\bq, \Omega)$  has one or two discontinuities, depending on the value of $\theta_q$. The real part of $\chi_L(\bq, \Omega)$  diverges logarithmically at these frequencies. Near $\Omega =2\Delta_0$, we found two logarithmic divergencies of $\mathrm{Im}\,\chi_L(\bq, \Omega)$, at $\Omega = 2\Delta_0 + v_F^2 |\bq|^2 \cos^2\theta_{\bq}/4\Delta_0$ and $\Omega = 2\Delta_0 + v_F^2 |\bq|^2 \sin^2\theta_{\bq}/4\Delta_0$, which we interpret as two dispersing longitudinal modes.  The real part of $\chi_L(\bq, \Omega)$ jumps by a finite value at each of these two frequencies. At $\theta_q$ directed towards a hot spot ($\theta_q = n \pi/2, n=0-3$) one of these modes does not disperse with $q$ and remains at $2\Delta_0$.  

We hope that our findings open a new avenue in the studies of the interplay between electronic criticality and 
the dynamics of a superconducting state mediated by strongly anisotropic attractive interaction emerging as the combination of quantum criticality and quantum geometry.

\section{Acknowledgments}
We acknowledge with thanks useful discussions with  D. Agterberg,  M. Dzero,  T. Hanaguri, and  T. Shibauchi.
The  work  was supported by U.S. Department
of Energy, Office of Science, Basic Energy Sciences,
under Award No. DE-SC0014402. K.R.I acknowledges support
from the Doctoral Dissertation Fellowship by the University of
Minnesota. K.R.I and A.V.C  acknowledge support from the Simons Foundation  Grant No. SFI-MPS-NFS-00006741-02 for the Simons Collaboration on New Frontiers in Superconductivity.

\section{Data availability}
The data that support the findings of this paper are available from the authors upon reasonable request.

\begin{appendix}
\section{Collective modes for a conventional s-wave superconductor}
\label{Appendix_A}

Collective modes of conventional BCS s-wave has been discussed in several references \cite{anderson1958random,anderson1958coherent,bardasis1961excitons,Anderson1958b,NNB1958, AndersonGauge, Volkov1975,SchmidSchon1975, ASchmid,VolkovKogan1973,Kulik1981, combescot2006collective,littlewood1982amplitude,VarmaLit1,maiti2013s+,maiti2015collective,phan2023following,podolsky2011visibility,althuser2025collective,schwarz2020classification,shimano2020higgs,udina2019theory}. In this section, we mostly re-derive them using diagrammatic technique. \\
We consider a one band system, captured by the Hamiltonian
\begin{align}
H_s=\sum_{\bk,\sigma} \xi(\bk) c^\dagger_{\bk,\sigma} c_{\bk,\sigma}-
    \sum_{\bk,\bp,\bq} V_0 c^\dagger_{\bk+\bq/2,\uparrow} c^\dagger_{-\bk+\bq/2 ,\downarrow} c_{-\bp+\bq/2 ,\downarrow}c_{\bp+\bq/2,\uparrow},
    \label{Hamiltonian for s-d}.
\end{align}
 where $V_0>0$. Within the mean field theory, at zero temperature, Eq.~\eqref{Hamiltonian for s-d} yields a homogeneous BCS s-wave superconductor ($\bq=0$) such that the  gap amplitude $\Delta_0$ is given by the following non-linear gap equation,
  \begin{align}
     1=V_0 \, N_0\int_{-\Lambda}^\Lambda d\xi_\bk \int_0^{2\pi} \dfrac{1}{2\,\sqrt{\xi^2_\bk+\Delta^2_0}}.
     \label{Gap equation s-d}
 \end{align}
We define the pair-pair susceptibility $\chi^s(\bq,\Omega_m)$  same as \eqref{chi_expression}. Within the RPA approximation, it is represented by the same set of diagrams from Fig.\ref{fig:vertex_figure} and gets the similar expression as defined in Eq.~\eqref{chi_ecpression_2}, 
\begin{align}
    \chi^s(\bq, \Omega_m)&=\int_k  G_s(k+\dfrac{q}{2})\, G_s(-k+\dfrac{q}{2})\, \Gamma^s_q- \int_k F_s(k+\dfrac{q}{2})\, F_s(-k+\dfrac{q}{2})\, \bar{\Gamma}^s_q,
    \label{chi_ecpression_2_s}
\end{align}
where the Green's functions are
\begin{align}
    G_s(k)=\dfrac{i\, \omega_m+\xi_\bk}{(i\, \omega_m)^2-E^2_{s,\bk}},\quad  F_s(k)=\dfrac{\Delta_0}{(i\, \omega_m)^2-E^2_{s,\bk}}, 
    \label{Green's function s-d}
\end{align}
and $E_{s,\bk}=\sqrt{\xi^2_\bk+\Delta_0^2}$ is the quasi-particle excitation energy, and the two-particle vertices  $\Gamma^s_q,\bar{\Gamma}^s_q$ are independent of internal momentum $\bk$, and the integration sign stands for $\int_k=T\sum_{\omega_\bk}\int d^2\bk/(2\pi)^2$. The equations for $\Gamma^s_q,\bar{\Gamma}^s_q$ are same as depicted in 
Fig.~\ref{fig:vertex_figure} b,c with the only exception that now the interaction line is $V_0$, carrying no momentum dependence. This gives the following expressions:
\begin{align}
    \Gamma^s_q&=1+V_0\, \Gamma^s_q \int_k G(k+\dfrac{q}{2})G(-k+\dfrac{q}{2})-V_0\, \bar{\Gamma}^s_q \int_k F(k+\dfrac{q}{2})F(-k+\dfrac{q}{2}),\\
     \bar{\Gamma}_q&=V_0\,\bar{\Gamma}^s_q\, \int_k G(k+\dfrac{q}{2})G(-k+\dfrac{q}{2}) - V_0\,\Gamma^s_q\,\int_k F(k+\dfrac{q}{2})F(-k+\dfrac{q}{2}).
\end{align}
Solving these equations, we find  
\begin{align}
\label{Gamma_expression_s}
 \Gamma^s_q&=\dfrac{1}{2}\left[\dfrac{1}{1-g \mathbf{\Pi}^s_T(q)}+ \dfrac{1}{1-g\mathbf{\Pi}^s_L(q)}\right], \\
 \bar{\Gamma}_q&=-\dfrac{1}{2}\left[\dfrac{1}{1-g \mathbf{\Pi}^s_T( q)}- \dfrac{1}{1-g \mathbf{\Pi}^s_L(q)}\right], 
 \label{bar_Gamma_expression_s}
\end{align}
where the transverse and longitudinal polarization bubble, $\mathbf{\Pi}^s_T$ and $\mathbf{\Pi}^s_L$, respectively, are defined as 
\begin{align}
    \mathbf{\Pi}^s_T(q)&=\int_k G_s(k+\dfrac{q}{2})G_s(-k+\dfrac{q}{2})+F_s(k+\dfrac{q}{2})F_s(-k+\dfrac{q}{2}) ,\label{piT for s-d}\\
  \mathbf{\Pi}^s_L(q)&=\int_k   \left( G_s(k+\dfrac{q}{2})G_s(-k+\dfrac{q}{2})-F_s(k+\dfrac{q}{2})F_s(-k+\dfrac{q}{2})\right).\label{piL for s-d}
\end{align}
We plug Eqs.~\eqref{Gamma_expression}-\eqref{bar_Gamma_expression_s} into the expression for the pair susceptibility $\chi^s(q)$, perform the analytic continuation $i\,\Omega_m \rightarrow \Omega+i\, \delta$, and get 
\begin{align}
    \chi^s(\bq,\Omega)=\dfrac{1}{2}\left[\dfrac{\Pi^s_T(q)}{1-V_0\, \Pi^s_T(q)}+\dfrac{\Pi^s_L(q)}{1-V_0\, \Pi^s_T(q)}\right],
    \label{Suscep_Eq_S_wave}
\end{align}

where $\Pi^s_{T,L}(q)=\Pi^s_{T,L}(\bq,\Omega)=\mathbf{\Pi}^s_{T,L}(\bq,i\, \Omega_m\rightarrow \Omega+i\, \delta)$. We integrate over the Matsubara frequency in the expression Eq.~\eqref{piT for s-d}-\eqref{piL for s-d}, and then  take the analytic limit $\Omega_m \rightarrow \Omega+i\, \delta$, which gives ,
\begin{align}
    \Pi^s_T(q)=\dfrac{N_0}{4}\int_{-\Lambda}^{\Lambda} d\xi_\bk \int_0^{2\pi}\dfrac{d\theta_\bk}{2\pi}  \left[1+\dfrac{\xi_{\bk+\bq/2}\xi_{\bk-\bq/2}+ \Delta^2_0}{E_{s,\bk+\bq/2}E_{s,\bk-\bq/2}}\right] & \left[\dfrac{1}{E_{s,\bk+\bq/2}+E_{s,\bk-\bq/2}-(\Omega+i \delta)} \right. \nn & \left. +\dfrac{1}{E_{s,\bk+\bq/2}+E_{s,\bk-\bq/2}+(\Omega+i \delta)}\right] \label{transverse bubble for s-wave}
    \end{align}
 \begin{align}
    \Pi^s_L(q)=\dfrac{N_0}{4}\int_{-\Lambda}^{\Lambda} d\xi_\bk \int_0^{2\pi}\dfrac{d\theta_\bk}{2\pi}  \left[1+\dfrac{\xi_{\bk+\bq/2}\xi_{\bk-\bq/2}-\Delta^2_0}{E_{s,\bk+\bq/2}E_{s,\bk-\bq/2}}\right] &\left[\dfrac{1}{E_{s,\bk+\bq/2}+E_{s,\bk-\bq/2}-(\Omega+i \delta)} \right. \nn & \left. +\dfrac{1}{E_{s,\bk+\bq/2}+E_{s,\bk-\bq/2}+(\Omega+i \delta)}\right] 
    \label{longitudinal bubble for s-wave}.
\end{align}
We note that the gap equation Eq.~\eqref{Gap equation s-d} can be recast in terms of transverse polarization bubble $1=V_0 \Pi^s_T(0)$. We use this relation to simplify Eq.~\eqref{Suscep_Eq_S_wave} to 
\begin{align}
    \chi^s_{T,L}(q)=-\dfrac{1}{2\, V_0} \left[\dfrac{1}{V_0\, \delta\Pi^s_{T,L}(q)}+1\right],
\end{align}
where $q=(\bq,\Omega)$ and $\delta\Pi^s_{T,L}(q)=\Pi^s_{T,L}(q)-\Pi^s_T(0)$. 
 To compare with the  corresponding expression used for NFMS, Eq.~\eqref{NFMS chi expression}, we write $\Pi^s_{T,L}(q)=N_0 \int \dfrac{d\theta_\bk}{2\pi} \Pi^s_{T,L}(\theta_\bk,q)$, where $\Pi^s_{T,L}(\theta_\bk,q)$ is the local polarization bubble -- the one in Eqs.~
 \eqref{transverse bubble for s-wave},\eqref{longitudinal bubble for s-wave} integrated only over the $\xi_\bk$,
 and define $g=N_0\, V_0$. In these notations,  
\begin{align}
  \chi^s_{T,L}(q)=-\dfrac{N_0}{2\, g} \left[\dfrac{1}{g\, \int \dfrac{d\theta_\bk}{2\pi} \delta\Pi^s_{T,L}(\theta_\bk,q)}+1\right].   
\end{align}
It has the same structure as  Eq.~\eqref{NFMS chi expression} for NFMS, but with $\Delta_0$ instead of $\Delta (\theta_k)$ \eqref{Gap form}. 
\subsection{Transverse Susceptibility}
\label{Appendix_A_transverse}
We call the transverse part of the pair-pair susceptibility, 
\begin{align}
    \chi^s_{T}(q)=-\dfrac{1}{2\, V_0} \left[\dfrac{1}{V_0\, \delta\Pi^s_{T}(q)}+1\right],
\end{align}
where $\delta\Pi^s_T(q)=\Pi^s_T(q)-\Pi^s_T(0)$, and has the following expression using Eq.~\eqref{transverse bubble for s-wave},
\begin{align}
    \delta \Pi^s_T(q)&=-\dfrac{1}{4}\int_0^{2\pi} \dfrac{d\theta_\bk}{2\pi} \int_{-\infty}^\infty d\xi_\bk  \dfrac{E_{s,\bk+\bq/2}+ E_{s,\bk-\bq/2}}{E_{s,\bk+\bq/2}\, E_{s,\bk-\bq/2}} \dfrac{\Omega^2-\left(\xi_{\bk+\bq/2}-\xi_{\bk-\bq/2}\right)^2}{(\Omega+i \delta)^2-\left(E_{s,\bk+\bq/2}+E_{s,\bk-\bq/2}\right)^2}.
     \label{tranverse susceptibility eq s-d}
\end{align}
As a result, the dispersion of the collective mode is determined from \begin{align}
    \delta\Pi^s_T(\bq,\Omega)=0
\end{align}
To study $\chi^s_T(q)$ at small momentum $|\bq|\ll k_F$ and small frequency $\Omega \ll E_F$ regime, we linearize the fermionic dispersion in $|\bq|$: $\xi_{\bk\pm\bq/2}=\xi_\bk\pm v_F |\bq| \cos(\theta_\bk-\theta_\bq)/2$, and perform the double expansion for $\delta\Pi^s_T(\bq,\Omega)$ in $v_F|q|/\Delta_0$ and $\Omega/\Delta_0$ up-to the second order (we assume $\Delta_0<< E_F$). After integrating over $\xi_\bk$, we get 
\begin{align}
    \delta\Pi^s_T(q)=-\int_0^{2\pi} \dfrac{d\theta_\bk}{2\pi}  \dfrac{(\Omega+i\,\delta)^2-v^2_F |\bq|^2 \cos^2(\theta_\bk-\theta_\bq)}{\Delta_0^2}=-\dfrac{(\Omega+i\delta)^2-v^2_F|\bq|^2/2}{\Delta_0^2}
    \label{delta_piT_for_s}
\end{align}
and  gives the dispersion of the Goldstone mode, 
\begin{align}
    \Omega(\bq)=v_F\, |\bq|/\sqrt{2}.
\end{align}
\subsection{Longitudinal Susceptibility }
\label{Appendix:Longtitudinal_for_s_wave}
We call the longitudinal part of the pair-pair susceptibility, 
\begin{align}
    \chi^s_{L}(q)=-\dfrac{1}{2\, V_0} \left[\dfrac{1}{V_0\, \delta\Pi^s_{L}(q)}+1\right],
\end{align}
where $\Pi^s_L(q)=\Pi^s_T(0)+\delta \Pi^s_L(q)$ and has the following expression using Eq.~\eqref{longitudinal bubble for s-wave},
\begin{align}
    \delta \Pi^s_L(q)&=-\dfrac{1}{4}\int_0^{2\pi}\dfrac{d\theta_\bk}{2\pi} \int_{-\infty}^\infty d\xi_\bk  \dfrac{E_{s,\bk+\bq/2}+ E_{s,\bk-\bq/2}}{E_{s,\bk+\bq/2}\, E_{s,\bk-\bq/2}} \dfrac{(\Omega)^2-4 \Delta_0^2-\left(\xi_{\bk+\bq/2}-\xi_{\bk-\bq/2}\right)^2}{(\Omega+i\delta)^2-\left(E_{s,\bk+\bq/2}+E_{s,\bk-\bq/2}\right)^2}.
     \label{delta_piL_equation_s}
\end{align}
For small $\bq$, we linearize the fermionic dispersion such that $(\xi_{\bk+\bq/2}-\xi_{\bk-\bq/2})^2\approx v^2_F|\bq|^2 \cos^2(\theta_\bk-\theta_\bq)$ in the numerator, and $(E_{s,\bk+\bq/2}-E_{s,\bk-\bq/2})^2 \approx 4 \Delta_0^2+4 \xi^2_\bk+v^2_F|\bq|^2 \cos^2(\theta_\bk-\theta_\bq)$  in the denominator of the integrand of Eq.~\eqref{delta_piL_equation_s}, while the rest of the integrand is non-singular and can be evaluated at $\bq=0$. After integrating over $\xi_\bk$,  we obtain 
\begin{align}
    \delta\Pi^s_L(q)=-\int_0^{\pi} \dfrac{d\theta_\bk}{\pi} \dfrac{\sec^{-1}\left(\dfrac{2 \Delta_0}{\sqrt{4 \Delta^2_{0} +v^2_F|\bq|^2 \cos^2\theta_\bk-(\Omega+i \delta)^2}}\right) \sqrt{4 \Delta^2_{0} +v^2_F|\bq|^2 \cos^2\theta_\bk -(\Omega+i \delta)^2}}{\sqrt{(\Omega+i \delta)^2-v^2_F|\bq|^2 \cos^2\theta_\bk}}.
    \label{delta piL for s}
\end{align}
\begin{figure}[]
  \centering
  \subfigure[]{\includegraphics[width = 0.45 \textwidth]{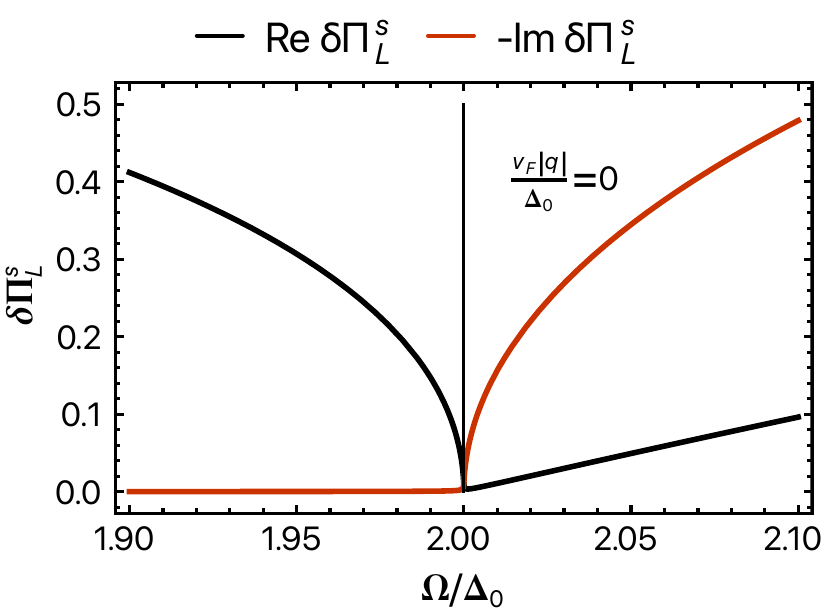}} \hspace{1 cm}
 \subfigure[]{\includegraphics[width = 0.45 \textwidth]{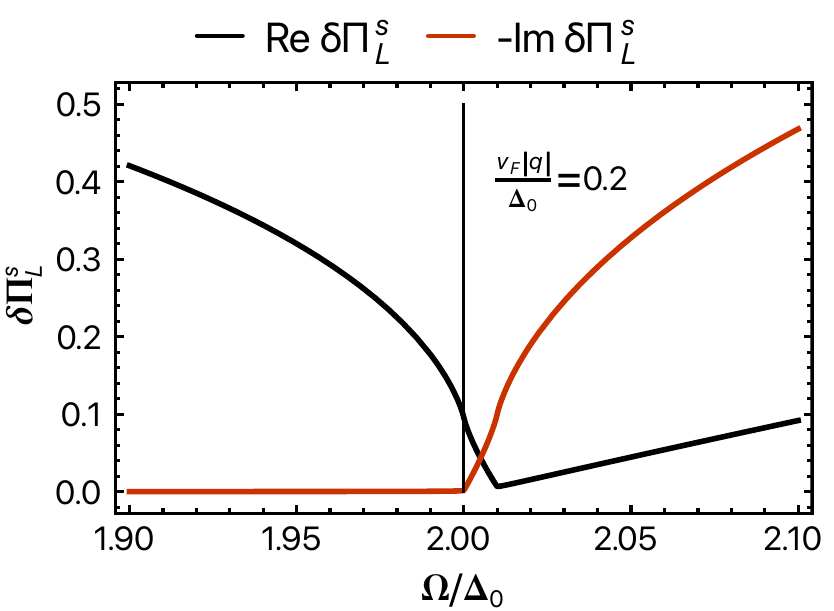}}
 \caption{Frequency dependence of the real and imaginary parts of the retarded longitudinal polarization bubble, $\delta\Pi^s_L(\bq,\Omega)$ for (a) $\bq=0$ and (b) $v_F |\bq|/\Delta_0=0.2$ for an s-wave superconductor with aconstant gap $\Delta_0$.}
\label{deltaPiL for swave}
 \end{figure} 
\begin{figure}[]
    \centering
    \includegraphics[width=0.5\linewidth]{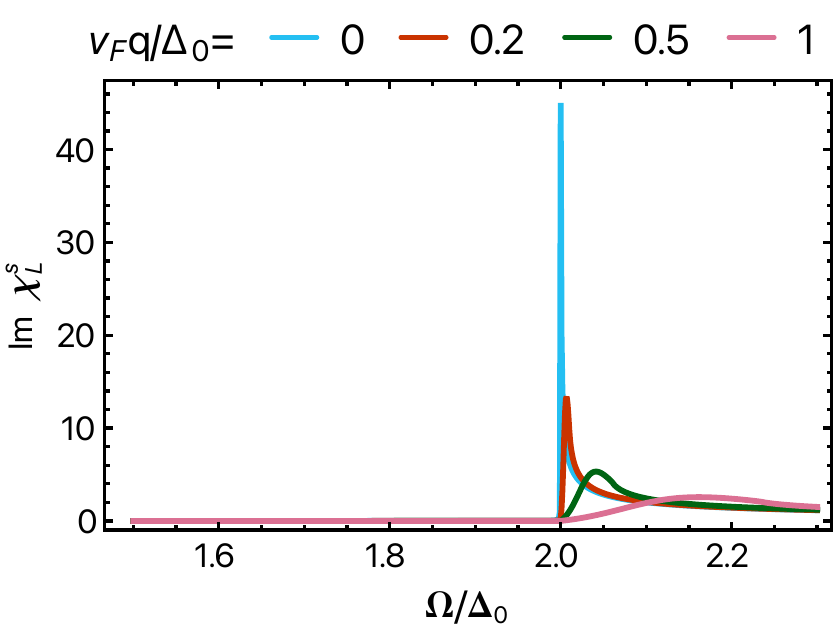}
    \caption{Frequency  dependence of the imaginary part of retarded longitudinal pair-pair susceptibility $\chi^s_L(|\bq|,\Omega)$ for a set of values of $|\bq|$ for an s-wave superconductor with a constant gap $\Delta_0$.}
    \label{chiL for swave}
\end{figure}
 
We note that $\theta_\bq$, direction of  momentum $\bq$ does not appear in the integrand of Eq.~\eqref{delta piL for s}. This is due to the isotropic nature of the  gap. As a  result, $\delta\Pi^s_L(q)$ is isotropic. For $\bq=0$, \begin{align}\delta\Pi^s_L(0,\Omega)= - \sec^{-1}\left(\dfrac{2 \Delta_0 }{\sqrt{4 \Delta^2_{0}  -(\Omega+i \delta)^2}}\right) \dfrac{\sqrt{4\Delta^2_{0}-(\Omega+i \delta)^2}}{\Omega}.\end{align}  We plot the real and imaginary parts of $\delta\Pi^s_L(0,\Omega)$ as a function of $\Omega$ in Fig.\ref{deltaPiL for swave}a.  Near $\Omega=2\Delta_0$, we expand in $\epsilon=\Omega-2\Delta_0$ and find 
\begin{align}
    \delta\Pi^s_L(0,\Bar{\epsilon})\propto -I \dfrac{\pi}{2}\sqrt{\Bar{\epsilon}}+\Bar{\epsilon},
\end{align} where $\Bar{\epsilon}=\epsilon/\Delta_0$ For $\epsilon<0$, $\delta\Pi^s_L$ is purely real, while for $\epsilon>0$, it has both a real part $\propto \epsilon$ and an imaginary part $\propto \sqrt{\epsilon}$. The real part vanishes at $\Omega=2\Delta_0 (\epsilon=0)$. This sets Im $\chi^s_L(0,\Omega)\propto -\text{Im}\delta\Pi^s_L/\left[(\text{Im}\delta\Pi^s_L)^2+(\text{Re}\delta\Pi^s_L)^2\right]$ vanishes for $\epsilon<0$ and diverges as $1/\sqrt{\epsilon}$   for $\epsilon>0$. This is shown in Fig.\ref{chiL for swave} (blue curve). The peak is commonly called a Higgs peak at energy $\Omega^s_\text{Higgs}=2\Delta_0$. For finite $\bq$, we numerically integrate Eq.~\eqref{delta piL for s} and plot the real and imaginary parts of $\delta\Pi^s_L(\bq,\Omega)$ in Fig.\ref{deltaPiL for swave}(b) and the corresponding Im $\chi^s_L(\bq,\Omega)$ in Fig.\ref{chiL for swave}.  To understand the numerical results near $\Omega=2\Delta_0+\epsilon$, we  assume and then verify that  the dominant  contribution to $\delta\Pi^s_L(\bq,\Omega)$ comes from  $\theta_\bk \approx 0$  and $\theta_\bk \approx \pi/2$ and approximate 
$\cos\theta_\bk \approx 1-\delta \theta^2_\bk$ near $\theta_\bk=0$ and $\approx -\delta\theta_\bk$ near $\theta_\bk=\pi/2$. We then obtain 
\begin{align}
    \delta\Pi^s_L(\bq,\Omega)\approx  \dfrac{\pi}{2}\int_0^{\Bar{\xi}_0/\Bar{|\bq|}} d\delta\theta_\bk \sqrt{\Bar{\xi}_0^2-\Bar{|\bq|}^2 \delta\theta^2_\bk}-I \dfrac{\pi}{2} \int_0^{\Bar{\xi}_1/\Bar{|\bq|}} d\delta\theta_\bk  \sqrt{\Bar{\xi}_1^2-\Bar{|\bq|}^2 \theta^2_\bk} \approx \dfrac{\pi}{2\, \Bar{|\bq|}} \left(\Bar{\xi}_0^2-I \Bar{\xi}_1^2 \right)
    \label{deltaPiL for s qnot0}
\end{align}
where  $\Bar{|\bq|}= v_F |\bq|/2 \Delta_0$, $\Bar{\xi}^2_0= \left(4 \Delta^2_0+v^2_F |\bq|^2-\Omega^2\right)/4 \Delta^2_0=\Bar{|\bq|}^2-\Bar{\epsilon}-\Bar{\epsilon}^2/4$ and  $\Bar{\xi}^2_1= -\left(4 \Delta^2_0-\Omega^2\right)/4 \Delta^2_0= \Bar{\epsilon}+\Bar{\epsilon}^2/4$. Using the expression \eqref{deltaPiL for s qnot0}, we find 
\begin{align}
    \text{Im}\chi^s_L(\bq,\epsilon)\propto \dfrac{\left[\Bar{\xi}_1(\epsilon)\right]^2}{\left(\Bar{|\bq|}^2-\left[\Bar{\xi}_1(\epsilon)\right]^2\right)+\left[\Bar{\xi}_1(\epsilon)\right]^4}.
\end{align}  and has a peak at a position $ \Bar{|\bq|}^2=\left[\Bar{\xi}_1(\epsilon)\right]^2 \rightarrow$ $\Omega^s_\text{peak}(\bq)=\sqrt{4 \Delta^2_0+v^2_F |\bq|^2}$.  

\section{Correction to the transverse susceptibility  from the  cold  regions}
\label{Appendix_B}
In this section we discuss a correction to the expression of transverse susceptibility, $\chi_T(\bq,\Omega)$, derived in Sec.\ref{Sec: Phase mode}, Eq.~\eqref{simplified tranverse susceptibility} which we get when we consider the effect of cold spot regions on it. We start with the expression for $\delta\Pi_T(\theta_\bk,\bq,\Omega)$, defined in Eq.~\eqref{delta piT equation}, which after integration over $\xi_\bk$ takes the following expression (we take $\theta_\bq=0$ for simplicity), 
\begin{align}
    \delta\Pi_T(\theta_\bk,\bq,\Omega)=\sec^{-1}\left(\dfrac{2\Delta(\theta_\bk)}{\sqrt{4 \Delta^2(\theta_\bk)+v^2_F|\bq|^2\cos^2\theta_\bk-\Omega^2}}\right) \dfrac{\sqrt{\Omega^2-v^2_F|\bq|^2 \cos^2\theta_\bk}}{\sqrt{4 \Delta^2(\theta_\bk)+v^2_F|\bq|^2\cos^2\theta_\bk-\Omega^2}},
    \label{delta pi eq1}
\end{align}
where $\Delta(\theta_\bk)=\Delta_0 \exp(-2\tan^22\theta_\bk/g)$. In Sec.\ref{Sec: Phase mode}, we expanded Eq.\eqref{delta piT equation} in small $\Omega/\Delta(\theta_\bk)$ and $v_F |\bq|/\Delta(\theta_\bk)$ and find $\delta\Pi_T(\theta_\bk,\bq,\Omega)=(\Omega^2-v^2_F|\bq|^2 \cos^2\theta_\bk)/4\Delta^2(\theta)$. This is correct as long as $\theta_\bk$ is not very close to the cold spots where the gap is exponentially small. When $\theta_\bk$ is close to cold spots, we expand Eq.~\eqref{delta pi eq1} near the cold spot where $\Delta(\theta_\bk)\approx \Delta_0 \exp(-1/4 g\delta\theta^2_\bk)$, $\cos^2\theta_\bk\approx 1/2$, and $\delta\theta_\bk=\theta_\bk-\theta_c$, $\theta_c=\pi/4$. This gives,  
\begin{align}
    \delta\Pi_T(\theta_\bk,\bq,\Omega)=-\dfrac{1}{2}\log \dfrac{\left|\Omega^2-\dfrac{v^2_f|\bq|^2}{2}\right|}{\Delta^2_0}-\dfrac{1}{4 g \delta\theta^2_\bk}+ i\, \dfrac{\pi}{2}\,\Theta\left(\Omega^2-\dfrac{v^2_f|\bq|^2}{2}\right),
\end{align}
where $\Theta(x)$ is Heaviside-Theta function. We put this expression into Eq.~\eqref{delta piT equation} and integrate near the cold spot region up-to a width $\delta\theta_0$ such that $\Omega^2-v^2_F|\bq|^2/2 \sim \Delta^2_0 \exp(-1/2 g\delta\theta^2_0) $, $\cos^22\theta_\bk\sim 4 \delta\theta^2_\bk$ and find 
\begin{align}
    \chi_T\propto& \int_0^{\delta\theta_0} d\delta\theta_\bk \dfrac{1}{4 g \delta\theta^2_\bk}\left[\dfrac{1}{4g\delta\theta^2_\bk \left(-\dfrac{1}{2}\log \dfrac{\left|\Omega^2-\dfrac{v^2_f|\bq|^2}{2}\right|}{\Delta^2_0}-\dfrac{1}{4 g \delta\theta^2_\bk}+ i\, \dfrac{\pi}{2}\,\Theta\left(\Omega^2-\dfrac{v^2_f|\bq|^2}{2}\right)\right)}+1\right]\\
    & \propto \int_0^{\delta\theta_0} d\delta\theta_\bk \dfrac{1}{4g\delta\theta^2_\bk}\left[-1-4 g\delta\theta^2_\bk \left(\dfrac{1}{2}\log \dfrac{\Delta^2_0}{\left|\Omega^2-\dfrac{v^2_f|\bq|^2}{2}\right|}+ i\, \dfrac{\pi}{2}\,\Theta\left(\Omega^2-\dfrac{v^2_f|\bq|^2}{2}\right)\right)+1\right]\\
    &\propto  \left(\log \dfrac{\Delta^2_0}{\left|\Omega^2-\dfrac{v^2_f|\bq|^2}{2}\right|}\right)^{1/2}+ i\, \dfrac{\pi}{2}\,\Theta\left(\Omega^2-\dfrac{v^2_f|\bq|^2}{2}\right) \left(\log \dfrac{\Delta^2_0}{\left|\Omega^2-\dfrac{v^2_f|\bq|^2}{2}\right|}\right)^{-1/2}, 
\end{align}
where we used $\delta\theta_0=\left(2 g \log \dfrac{\Delta^2_0}{\left|\Omega^2-\dfrac{v^2_f|\bq|^2}{2}\right|}\right)^{-1/2}$.  
\section{Width and height of the peaks of transverse susceptibility, $\chi_T$}
\label{Appendix_C}
In this section we derive the height and width of the peaks of Im$\chi_T(\bq,\Omega)$ presented in the Sec.\ref{Sec: Phase mode}. We rewrite the expression for Im$\chi_T(\bq,\Omega)$ , 
\begin{align}
    \text{Im}\chi_T(|\bq|,\theta_\bq,\Omega)&=\dfrac{N_0}{g^2}\dfrac{\Theta\left(1-\dfrac{|\Omega|}{v_F|\bq|}\right)}{\Omega \sqrt{v^2_F|\bq|^2-\Omega^2}}\left[\psi\left(\theta_\bq+\cos^{-1}\dfrac{\Omega}{v_F|\bq|}\right)+\psi\left(\theta_\bq-\cos^{-1}\dfrac{\Omega}{v_F|\bq|}\right) \right],
    \label{chi T expression 2}
\end{align}
where $\psi(\theta)=\Delta^2(\theta)/\cos^42\theta$ and $\Delta(\theta)=\Delta_0\, \exp^{-\tan^22\theta/g}$. $\psi(\theta)$ is a highly peaked function of $\theta$, and its maxima occurs when $\theta=n \pi/2,. n=0-3$ with a magnitude of $\Delta^2_0$. As a result, the peak of Im$\chi_T(|\bq|,\theta_\bq,\Omega)$ appears when \begin{align}
    \theta_\bq\pm \cos^{-1}\dfrac{\Omega}{v_F |\bq|}=n \dfrac{\pi}{2}, \quad n=0-3
\end{align}
This corresponds to two unique solution: $\Omega_1=v_F |\bq|\, \cos\theta_\bq$ and  $\Omega_2=v_F |\bq|\, \sin\theta_\bq$. The height of these peaks are equal to
\begin{align}
 h_s=\text{Im}\chi_T(|\bq|,\theta_\bq,\Omega_s)=\dfrac{2\, N_0\, \Delta^2_0}{g^2 v^2_F |\bq|^2\, \sin2\theta_\bq} , \quad i=1,2
\end{align}
We define the width of these peaks as width at half maxima such that
\begin{align}
   \dfrac{\psi\left(\theta_\bq\pm \cos^{-1}\dfrac{\Omega}{v_F |\bq|}\right)}{\Omega \sqrt{v^2_F|\bq|^2-\Omega^2}}= \dfrac{\Delta^2_0}{ v^2_F |\bq|^2\, \sin2\theta_\bq}.
   \label{half width}
\end{align}
We define $x=\Omega/v_F|\bq|$ and write Eq.~\eqref{half width} using the expression for $\psi$. This gives 
\begin{align}
    \dfrac{\exp^{-2\tan^2(2\theta_\bq\pm 2\cos^{-1}x)/g}}{\cos^4(2\theta_\bq\pm2\cos^{-1}x)}=\dfrac{x \sqrt{1-x^2}}{\sin2\theta_\bq}
    \label{full exp for width}
\end{align}
 We first focus on the peak at $x_0=\Omega_1/v_F |\bq|\,= \cos\theta_\bq$. For this peak, we write $x=x_0+\delta x$, and approximate $\cos^{-1}x\approx  \cos^{-1}x_0-\delta x/\sin\theta_\bq$, $\tan^2(2\theta_\bq-2\cos^{-1}x)\approx  \dfrac{4(\delta x)^2}{\sin^2\theta_\bq}$, and $\cos^4(2\theta_\bq-2\cos^{-1}x)\approx 1$ and $x\sqrt{1-x^2}\approx \sin2\theta_\bq/2$. Combining all the factors we get, $ \exp (-8 (\delta x)^2/ g \sin^2\theta_\bq) =1/2$. This gives the width proportional to 
 \begin{align}
   \delta x\propto |\sin\theta_\bq|.
 \end{align}
 This shows that at small $\theta_\bq$, the peak at $\Omega_1=v_F |\bq|\, \cos\theta_\bq$ is very sharp.
 On the other hand, for the peak at $\Omega_2=v_f |\bq|\sin\theta_\bq$, similar analysis will produce width $\delta x\propto \cos\theta_\bq$.
\section{Kramers Kronig Relation}
\label{Appendix_D}
 Let $\chi(\omega)=\chi_1(\omega)+I\, \chi_2(\omega)$ is a complex function of a complex variable $\omega$, where $\chi_{1,2}(\omega)$ are real function. Suppose $\chi(\omega)$ is analytic on the upper half plane of $\omega$ and follow the Kramers-Kronig relation
\begin{align}
    \chi_1(\Omega)&=\dfrac{1}{\pi} P\int_{-\infty}^{\infty} d\omega' \dfrac{\chi_2(\omega')}{\omega'-\omega}, \label{im to re}\\
    \chi_2(\Omega)&=-\dfrac{1}{\pi} P\int_{-\infty}^{\infty} d\omega' \dfrac{\chi_1(\omega')}{\omega'-\omega} \label{re to im},
\end{align} where $P$ stands for the principle value of the integration. We prove that if one of the components of $\chi(\omega)$ (either real or imaginary) has a logarithmic singularity at some $\omega=\omega_0$, the other component will show a discontinuity at $\omega_0$. Since Eq.\eqref{re to im} and \eqref{im to re} have the some functional structure (except the presence the minus sign), irrespective of which component of $\chi(\omega)$ has the logarithmic singularity, the other will always have a discontinuity. Let's consider \begin{align}
    \chi_2(\omega)=A \log|\omega-\omega_0|.
\end{align}  Then using Eq.~\eqref{im to re} we find, 
\begin{align}
    \chi_1(\omega)=\dfrac{A}{\pi}P \int_{-\infty}^{\infty} d\omega'\dfrac{\log |\omega'-\omega_0|}{\omega'-\omega}
    \label{chi eq 1}
 \end{align}. We shift the variable $x=\omega'-\omega_0$, and define $\omega=\omega_0+\epsilon$, then Eq.\eqref{chi eq 1} becomes
 \begin{align}
     \chi_1(\epsilon)=\dfrac{A}{\pi} \int_{-\infty}^{\infty}\,dx \dfrac{\log |x|}{x-\epsilon}=\dfrac{2\epsilon\, A}{\pi} P \int_0^{\infty} \,dx \dfrac{\log x}{x^2-\epsilon^2}=\dfrac{2 A \pi}{sgn(\epsilon)}.
 \end{align} where $P\int_0^\infty dx\dfrac{\log x}{x^2-\epsilon^2}=\pi^2/4 |\epsilon|$. $sgn(x)$ is the signum function, it is $-1$ when $x<0$ and $1$ when $x>0$. As a reuslt, $\chi_1(\epsilon)$ undergoes a jump from $-2 A \pi$ to $2 A \pi$ across $\omega=\omega_0$.  \\
 On the other hand, if the real part has logarithmic singularity, $\chi_1(\omega)=A\, \log|\omega-\omega_0|$, using Eq.~\eqref{re to im}, one would find that $\chi_2(\omega)$ undergoes a jump from $2 A \pi$ to $-2 A \pi$ across $\omega=\omega_0$.
 \end{appendix}
 
\bibliography{biblio}

\end{document}